\documentclass[12pt]{iopart}
\usepackage{graphicx}

\newcommand{\rref}[1]{reference~\cite{#1}}

\newcommand{\rsref}[1]{references~\cite{#1}}
\newcommand{\Rsref}[1]{References~\cite{#1}}
\newcommand{\phig}{\phi_{\rm g}}
\newcommand{\tw}{t_{\rm w}}
\newcommand{\cI}{c_I(t,\tau)}

\begin{document}

\review[Slow dynamics] {Slow dynamics in glassy soft matter}

\author{Luca Cipelletti\footnote[3]{To whom correspondence should be addressed
(lucacip@gdpc.univ-montp2.fr)} and Laurence Ramos}

\address{LCVN UMR 5587, Universit\'{e} Montpellier II and CNRS, P. Bataillon, 34095 Montpellier, France}

\begin{abstract}

Measuring, characterizing and modelling the slow dynamics of
glassy soft matter is a great challenge, with an impact that
ranges from industrial applications to fundamental issues in
modern statistical physics, such as the glass transition and the
description of out-of-equilibrium systems. Although our
understanding of these phenomena is still far from complete,
recent simulations and novel theoretical approaches and
experimental methods have shed new light on the dynamics of soft
glassy materials. In this paper, we review the work of the last
few years, with an emphasis on experiments in four distinct and
yet related areas: the existence of two different glass states
(attractive and repulsive), the dynamics of systems very far from
equilibrium, the effect of an external perturbation on glassy
materials, and dynamical heterogeneity.

\end{abstract}


\submitto{\JPCM}


\tableofcontents

\section{Introduction} \label{sec:intro}

A wide range of soft materials exhibit slow relaxations strongly
reminiscent of the glassy dynamics observed in hard condensed
matter. These materials include concentrated or strongly
interacting colloidal suspensions, emulsions, surfactant systems,
polymeric or colloidal gels, and foams \cite{Cates2000}. Most of
these materials are out of thermodynamical equilibrium: for
systems far from equilibrium, a continuous evolution of the
mechanical and dynamical properties is observed, in analogy with
the slowing down of the dynamics in aging hard glasses. Several
different terms are used in the literature to designate
out-of-equilibrium soft materials, depending on the context and
the authors' views: gels, glasses, and jammed \cite{LiuNature1998}
systems are but a few examples. Here, we will term them quite
generally ``soft glassy systems'' \cite{SollichPRL1997} or simply
soft glasses. Soft glasses are non-ergodic systems, since they are
able to explore only a restricted portion of the total phase
space. Experimentally, however, determining whether or not a given
system is non-ergodic may prove to be a difficult task, since only
a limited range of length and time scales can be probed. In the
following we will therefore use the term ``soft glassy system''
with a wider connotation, including also materials that are
ergodic but still exhibit very slow relaxations.

Soft glassy systems attract a wide interest. On the one hand, they
are ubiquitous in industrial applications, for example in the
food, cosmetic, paint, pharmaceutical, and oil recovery industry.
For most applications, understanding and controlling their
dynamical and rheological properties, as well as their long term
evolution and stability is of fundamental importance. On the other
hand, soft glasses are much studied at a more fundamental level.
For example, systems such as colloidal hard spheres are taken as
model systems for investigating the general behavior of glasses,
their relevant time and length scales being more easily accessible
than in molecular glasses. Moreover, soft systems allow one to
control in great detail and to significantly vary the interactions
between their constituents. Therefore, it is possible to
investigate how different interactions influence the dynamical
behavior of soft glassy materials and the generality of the
behavior observed for model systems can be tested.

Until very recently, the experimental characterization of the
dynamics was limited to relatively short time scales (e.g. up to a
few hundred seconds in light scattering experiments) and only the
average dynamics were accessible. Theoretically, very few
quantitative models were available, and their applicability was
restricted to a few model systems (e.g. the mode coupling theory,
MCT \cite{GotzeRepProgPhys}, applied to colloidal suspensions
interacting via a hard sphere potential). Recent advances include
progress both on the conceptual and experimental side. The concept
of jamming \cite{LiuNature1998} was proposed to describe in a
unified way ---albeit somehow qualitatively--- the ergodic to
non-ergodic transition in many systems, including molecular
glasses, granular materials, and colloids. On the other hand, the
MCT has been extended to include concentrated systems with
attractive interactions (for a review see reference
\cite{DawsonCOCIS2002}) and attempts have been made to extend its
validity to more diluted attractive systems \cite{KroyPRL2004},
thus bridging the gap between colloidal gels and glasses. Thus,
both the jamming picture and the MCT attempt to give a unified
vision of the ergodic to non-ergodic transition, the scope of the
former being wider and the latter providing more quantitative
predictions. Finally, dynamical heterogeneity has emerged as a key
feature of the slow dynamics of glassy systems
\cite{RichertJPCM2002} and many theories have been developed based
on or incorporating dynamical heterogeneity.

On the experimental side, recent advances include both the design
of new systems with interactions that can be fine-tuned and the
development of new methods and techniques to measure very slow and
heterogeneous relaxations in out-of-equilibrium systems.
Experimentally, controlling the transition to non-ergodicity
(possibly in a reversible way) is often quite difficult. As a
consequence, systems for which a transition from a fluid to a
solid state can be reversibly obtained are particularly
interesting. They include thermoresponsive systems such as
thermoreversible sticky hard spheres
\cite{SolomonPRE2001,PontoniJChemPhys2003} or emulsions
\cite{KohChemCommun2000,KohPhysChemChemPhys2002}, swellable
microgel particles \cite{WuPRL2003,StiegerLangmuir2004}, star
polymers \cite{KapnitosPRL2000,
StiakakisPRE2002,StiakakisLangmuir2003}, and surfactant systems
\cite{RamosPRL2001,RamosEPL2004,ChenPRE2002}.

The most popular techniques for probing the dynamics of soft
systems are probably optical microscopy and dynamic light
scattering: both of them have benefited of major improvements in
the last years. Time-resolved confocal microscopy, together with
digital image processing, allows the trajectories of thousands of
fluorescently dyed particles to be measured simultaneously
\cite{HabdasCOCIS2002}, thus providing detailed information on the
dynamics at a microscopic level. Motion on very long time scales
(up to several days) has been measured by bleaching a cubic volume
of the sample and by measuring the fluorescence recovery due to
the diffusion of the unbleached particles
\cite{SimeonovaFaradayDiscuss2003}.

In light scattering experiments, information on the sample
dynamics are obtained by measuring the time autocorrelation
function of the scattered intensity, $g_2(t)$. In the single
scattering regime \cite{Pecora}, the dynamic structure factor (or
intermediate scattering function) $f(q,t)$ is proportional to
$\sqrt{g_2(t)-1}$, where the intensity autocorrelation function is
measured at a scattering vector $\bf{q}$ and the proportionality
factor depends on the setup detection. The advent of high
brilliance coherent X-ray sources has made possible X-photon
correlation spectroscopy (XPCS) \cite{DiatCOCIS1998}, which
extends dynamic light scattering experiments to much higher values
of $q$. For highly multiply scattering samples, the diffusing wave
spectroscopy (DWS) \cite{DWSGeneral} formalism allows the particle
mean squared displacement to be extracted from $g_2(t)$, provided
that the dynamics be homogeneous.  Until recently, the
applicability of light scattering and XPCS techniques to soft
glasses has been severely limited by the need to average the
measured intensity autocorrelation function over a duration much
longer than the system relaxation time. This limitation has been
overcame by the so-called multispeckle technique
\cite{WongRSI1993,BartschJChemPhys1997,LucaRSI1999,ViasnoffRSI2002},
where a multielement detector is used (typically a CCD camera) and
the intensity autocorrelation function is averaged not only over
time, but also over distinct speckles corresponding to different
elements of the detector. This scheme allows one to drastically
reduce the duration of a measurement, thus making dynamic
scattering techniques suitable for systems with very slow or
non-stationary dynamics. A conceptually similar approach is
followed in the so-called interleaved method, where a point
detector is used and the sample is slowly rotated in order to
illuminate sequentially the detector with different speckles
\cite{MullerProgColloidPolymSci1996}; for each speckle the
intensity correlation function is calculated for time delays
multiple of the period of rotation of the sample and $g_2$ is
finally obtained by averaging over all speckles. A recent
implementation of this scheme, more simple and efficient, is
described in reference \cite{PhamRSI2004}. To conclude this brief
overview of recent advances in scattering methods, we remark that
most scattering experiments probe the average dynamics and hence
are not sensitive to dynamic heterogeneities. Information on
temporal heterogeneity of the dynamics can be extracted from
higher order correlation functions
\cite{LemieuxJOSAA1999,LemieuxAPPOPT2001} or using the Time
Resolved Correlation (TRC) method \cite{LucaJPCM2003}, as it will
be discussed in some detail in section \ref{sec:hetero}.

In this paper we will review the experimental work of the last few
years on soft glassy systems, with an emphasis on microscopy,
light scattering and, to a less extent, rheology experiments.
References will be made also to some of the numerical and
theoretical investigations related to experiments, as well as to
experiments on granular materials that share intriguing analogies
with soft glasses. The literature is very abundant, reflecting the
growing interest in this field. We have chosen to focus on four
areas that seem to us particularly interesting and promising.
\Sref{sec:MCT} deals with the observation of two types of
colloidal glasses (repulsive and attractive), and the success of
the MCT in modelling the slow dynamics of these systems. Systems
very far from equilibrium, where stress relaxation appears to play
a major role in the slow dynamics are reviewed in section
\ref{sec:outofequilibrium}. Section \ref{sec:perturb} discusses
the effect of a perturbation on a glassy system, both in the
linear and in the non-linear regime. Finally, dynamical
heterogeneities in the slow dynamics are addressed in section
\ref{sec:hetero}.

\section{Attractive and repulsive glasses: experiments and the mode-coupling theory}
\label{sec:MCT}

Hard spheres have been investigated as a model system for the
glass transition for many years. Experimentally, colloidal
particles such as poly(methyl methacrylate) (PMMA) spheres
sterically stabilized and suspended in an organic solvent have
been shown to behave as hard spheres, forming a glass phase when
the volume fraction, $\phi$, becomes larger than $\phig \approx
0.58$ \cite{PuseyNature1986,PuseyPRL1987}. On the other hand, the
mode-coupling theory (MCT) introduced by G\"{o}tze
\cite{GotzeRepProgPhys} provides detailed predictions on the
dynamics of hard spheres when approaching the glass transition.
For a supercooled liquid of hard spheres\footnote{For hard
spheres, the temperature quench that is used in molecular systems
to obtain a supercooled fluid would correspond to a rapid increase
of volume fraction, above the crystallization volume fraction
$\phi_{\rm cryst} = 0.494$ and below $\phig$. Although in some
experiments centrifugation is used to realize such a quench, in
most cases supercooled samples are prepared by shear melting.} the
MCT predicts a two-step decay of the dynamic structure factor
$f(q,t)$. The initial relaxation ($\beta$ relaxation) is due to
the rattling of the particles inside the cage formed by their
neighbors. The $\beta$ relaxation is followed by a plateau and a
second relaxation ($\alpha$ relaxation), due to cage-escape
processes. At the glass transition, the ideal MCT predicts that
the characteristic time of the $\alpha$ relaxation diverges and
thus that the plateau extends to all times, indicating a complete
dynamical arrest. Note however that on very long time scales
ergodicity may be restored by thermally activated processes
(hopping), and attempts have been done to include such processes
in the MCT \cite{GotzeRepProgPhys}. Good agreement has been found
between the MCT and dynamic light scattering measurements of the
dynamic structure factor of hard sphere colloidal suspensions (for
a review on experimental tests of the MCT, see
\cite{GotzeJPCM1999}), although the critical packing fraction
$\phig^{\rm MCT} = 0.525$ predicted by the MCT is smaller than the
experimentally measured glass transition volume fraction $\phig
\approx 0.58$. To overcome this discrepancy, the $relative$ volume
fractions $(\phi -\phig)/\phig$ and $(\phi -\phig^{\rm
MCT})/\phig^{\rm MCT}$ are often used when comparing experiments
and theoretical or numerical work \cite{vanMegenPRE1994}.

In hard spheres, the dramatic slowing down of the dynamics when
approaching $\phig$ is due to the cage effect: the motion of any
given particle is increasingly hindered by its neighbors as the
particles are packed more tightly. The glass transition is
therefore driven by the repulsive (excluded volume) interaction
between the spheres, and the arrested phase thus formed is termed
{\it repulsive} glass. Recent theoretical
\cite{BergenholtzPRE1999,FabbianPRE1999,DawsonPRE2001} and
experimental
\cite{EckertPRL2002,PhamScience2002,ChenScience2003,BhatiaLangmuir2002}
work has shown that the addition of short-ranged, low-energy
attractive interactions can lead, surprisingly, to the melting of
such a repulsive glass. This can be understood in the framework of
the cage effect: particle bonding due to attractions results in
the increase of the available free volume, thus loosening and
eventually opening the cage. If the strength of the attraction is
increased further, however, a new arrested phase is formed,
because the bonds are sufficiently long-lived to effectively
confine the particles. This arrested phase is referred to as an
$attractive$ glass, although the term gel is also found in the
literature. For a short general introduction to the topic of
attractive $vs$ repulsive glass, see \cite{SciortinoNatMater2002}.
A brief and clear review of the experimental work on this topic is
given in \cite{PoonMRS2004}, while \rref{DawsonCOCIS2002}
discusses recent theoretical advances. In the following, we will
briefly recall the main results of MCT and then focus on the
experimental work.

\begin{figure}
\begin{center}
\includegraphics{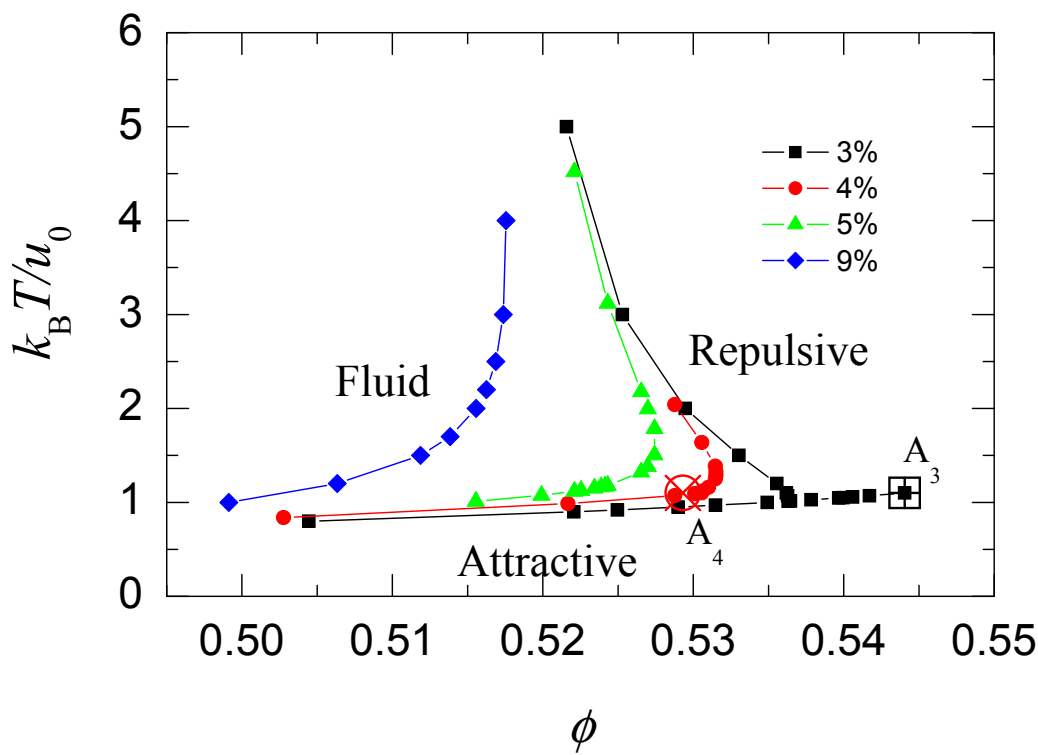}
\end{center}
\caption{\label{fig:2glasses} Theoretical phase diagram for hard
spheres of radius $R$ with a square well attractive interaction of
width $\Delta$ and depth $u_0$ (data from \rref{DawsonPRE2001}).
The different curves are labelled by the relative well width, $\xi
= \Delta/(2R+\Delta)$. Note the reentrant shape of the lines
separating the fluid region to the glassy regions for $\xi \leq
5\%$. For $\xi = 3\%$ a line separating the attractive and the
repulsive glass phases ends in the $A_3$ singular point. The $A_4$
singular point (circled cross) is estimated to be located on the
$\xi = 4.11\%$ line (not shown for clarity).}
\end{figure}

\Fref{fig:2glasses} shows the theoretical phase diagram for
particles interacting via a hard-core repulsion ($U(r) = \infty$
for $r < 2R$, where $U$ is the inter-particle potential and $R$ is
the particle radius) and a square well attraction ($U(r) = -u_0$
for $2R < r < 2R + \Delta$) \cite{DawsonPRE2001}. The different
curves are labelled by the relative width of the attractive well,
$\xi = \Delta/(2R+\Delta)$. Although the exact location of the
lines depends on the detailed shape of the potential and the
chosen approximated expression for the static structure factor,
the features shown in \fref{fig:2glasses} are quite general. The
lower part of the phase diagram corresponds to strong attractions
($u_0 \gg k_{\rm B}T$) and is occupied by the attractive glass.
The upper-right region is occupied by the repulsive glass, while
the upper-left part corresponds to the fluid phase. Note that at
small enough $\xi$ (typically less than about 0.1) the phase
diagram has a reentrant shape: there exists a range of volume
fractions for which, at constant $\phi$, a sample may be in the
fluid phase or in either the attractive or the repulsive glass
phase, depending on the strength of the attractions. For very
short-ranged attractions ($\xi $ less than about 0.04), at high
$\phi$ the attractive and repulsive glass regions are contiguous,
separated by a line ending in the so-called $A_3$ singular point
(open square with plus in \fref{fig:2glasses}). When $\xi$
increases, the line separating the attractive and the repulsive
glasses progressively shrinks, until the line and the $A_3$
singularity vanish and the so-called $A_4$ singularity is observed
(open circle with cross in \fref{fig:2glasses}). In the vicinity
of the $A_3$ and $A_4$ singular points, as well as at the
intersection of the attractive and repulsive glass lines, the
system is predicted to relax with a very broad distribution of
time scales, resulting in a logarithmic decay of the dynamic
structure factor $f(q,t)$ and a power law behavior of the particle
mean squared displacement
\cite{GotzePRE2002,SperlPRE2003,SperlPRE2004}. Simulations
performed by using either a square-well
\cite{FoffiPRE2002,ZaccarelliPRE2002,SciortinoPRL2003} or the
Asakura-Osawa potential
\cite{PuertasPRL2002,PuertasPRE2003,CatesJPCM2004}, have confirmed
the predictions of the MCT.

Experimentally, the more extensive tests of the two glasses
scenario have been performed by three independent groups. Pham and
coworkers \cite{PhamScience2002,PhamPRE2004} have used nearly
monodisperse, sterically stabilized PMMA particles to which
polystyrene, a non-adsorbing polymer, was added to induce a
short-ranged attraction via the depletion effect, leading to an
attraction well described by the Asakura-Osawa potential (for a
review on this much studied experimental system, see
\cite{PoonJPCM2002}). The strength of the attraction can be
controlled by changing the amount of the polymer, while the range
is of the order of the radius of gyration of the polymer. In
\rsref{PhamScience2002,PhamPRE2004} the phase diagram of this
system is established and both the structure and the dynamics of
the glass phases are studied in detail by means of light
scattering. The so called two-color technique
\cite{SegreJModOpt1995} is used to suppress multiple scattering,
and a recently introduced ``echo'' scheme \cite{PhamRSI2004} is
used to obtain properly ensemble-averaged measurements even when
dealing with non-ergodic samples. As expected from the MCT, at
fixed $\phi$ the non-ergodicity ---or Debye-Waller--- factor
$f(q,t \rightarrow \infty)$ for the attractive glass is found to
be higher than that for the repulsive glass, because the motion of
clustered particles is more restrained than that of repulsive
spheres, which can rattle in the cage formed by their neighbors.
Accordingly, the liquid-order peak in the static structure factor,
$S(q)$, shifts to larger $q$ vectors when going from the repulsive
to the attractive glass. Additionally, $S(q)$ is less sharp for
attractive glasses, since they are locally more disordered than
repulsive glasses. In the vicinity of the $A_3$ point, a
logarithmic decay of $f(q,t)$ is observed over more than three
decades in time. Beyond the $A_3$ point, where the MCT predicts
that the attractive and repulsive glasses should merge into a
single arrested phase, the non-ergodicity factor becomes
independent of attraction strength, suggesting that indeed the two
glasses phases are indistinguishable. However, quite intriguingly,
the short-time behavior of $f(q,t)$ remains different, a feature
also reported in \rref{ChenPRE2003}.

Similar detailed investigations, leading to comparable results,
are presented by Eckert and Bartsch and by Chen and coworkers.
Eckert and Bartsch study a concentrated suspension of microgel
particles, to which linear polystyrene is added to induce an
attraction via the depletion effect
\cite{EckertPRL2002,EckertFaradayDiscuss2003}. The microgel
particles are bidisperse so as to suppress crystallization. This
two-component system is mapped onto an effective (polydisperse)
one-component system of slightly soft spheres (the repulsive
barrier is modelled by an inverse power potential $U(r) \sim
r^{-35}$), and is studied by static and dynamic light scattering.
The multispeckle technique
\cite{BartschJChemPhys1997,EckertFaradayDiscuss2003} is used to
access the long time dynamics and for studying non-ergodic
samples. Chen and coworkers
\cite{ChenPRE2002,ChenScience2003,ChenPRE2003} investigate by
small angle neutron scattering (SANS) and dynamic light scattering
(DLS) a concentrated aqueous solution of L64, a triblock copolymer
of the Pluronics family. L64 is composed of two end sections of
polyethylene oxide (PEO) and a central section of polypropylene
oxide (PPO). Both PPO and PEO become increasingly hydrophobic when
the temperature $T$ is raised, but PPO does so at a faster rate.
As a consequence, when $T$ is increased micelles are formed with a
PPO core and a PEO corona. At high enough temperature, the PEO
corona becomes sticky because of hydrophobic interactions, thus
providing an attraction between the micelles. The experimental
control parameters for this system are $T$ and the block copolymer
concentration, $C$. For comparison with the MCT phase diagram, the
$T$-$C$ experimental phase diagram is mapped onto a $\phi$-$u_0$
diagram by fitting the static structure factor obtained by SANS to
that predicted for adhesive hard spheres. Note that the DLS data
are taken at a $q$ vector much smaller than that corresponding to
the peak of $S(q)$, in contrast to the work of Pham $et$ $al.$ and
Eckart and Bartsch.

Several features of the two glasses scenario predicted by the MCT
have been observed also in other systems.
\Rsref{BhatiaLangmuir2002,GrandjeanEurophysLett2004} report a
rheology, SANS and dynamic light scattering investigation of a
solution of soft spheres, made of diblock copolymer micelles. An
attraction can be induced by synthesizing diblock polymers with a
controlled amount of stickers (ethyl acrylate units). Two distinct
glass phases and a reentrant behavior are observed. The rheology
experiments show that the elastic modulus $G_0$ of the attractive
glass is larger than that of the repulsive glass, a feature
predicted by the MCT. Note, however, that for these soft objects
the difference in $G_0$ for attractive and repulsive glasses
(about a factor of 2) is much less marked than for attractive hard
spheres (more than a decade, see \cite{BergenholtzLangmuir2003}).
The melting of a glass of soft spheres (star polymers) and a
reentrant gelation upon the addition of a linear polymer has been
observed by Stiakakis and coworkers \cite{StiakakisPRL2002}.
Although the phenomenology is strongly reminiscent of that
observed in other systems with depletion interactions, the authors
point out that the melting is due to the decrease of the range of
the soft repulsion between stars upon addition of the linear
polymer, while the gelation results from bridging flocculation
induced by the long polymers chains. Finally, the existence of a
reentrant glass transition in Laponite suspensions has been
speculated, but not yet proved experimentally
\cite{TanakaPRE2004}.

Qualitatively, the experiments mentioned above strongly support
the MCT picture of attractive and repulsive glassy states and show
that this scenario is robust with respect to changes in the exact
shape of the potential. A quantitative comparison is however much
more difficult. Indeed, even for the experimental system closer to
the theoretical models, i.e. the hard spheres with depletion
interactions studied by Pham and coworkers, it is difficult to
exactly locate a sample on the theoretical phase diagram, because
of uncertainties in the parameters describing the potential and in
the relative volume fraction $(\phi -\phig)/\phig$ of a sample
\cite{PhamPRE2004}. A comparison of the experimental and
theoretical phase diagrams is given in
\rref{BergenholtzLangmuir2003}, showing that the MCT somehow
underestimates the extent of the fluid pocket in between the
attractive and repulsive glass regions.

In view of the success of the MCT in modelling suspensions of
attractive particles at high volume fraction, attempts have been
made to extend it to lower $\phi$, in order to describe as a glass
transition the ``weak gelation'' observed in attractive systems
where particle bonds are non-permanent. In the ``cluster
mode-coupling theory'' of Kroy and coworkers \cite{KroyPRL2004},
the MCT is applied to clusters of particles, rather than to the
particles themselves. This theory predicts the existence of a
cluster phase that may freeze if the effective cluster volume
fraction is large enough. Recent simulations
\cite{DelGadoEurophysLett2003,DelGadoPRE2004} and experiments
\cite{SegrePRL2001} are consistent with this picture. A more
detailed discussion of the delicate interplay between aggregation,
dynamical arrest and phase separation that arises in relatively
low $\phi$ suspensions is given in \cite{CatesJPCM2004}. For a
discussion of colloidal gelation in connection to the glass
transition, see also the review by Trappe and Sandk\"{u}hler
\cite{TrappeCOCIS2004}, which focuses on the elastic properties of
these materials.

In spite of the large amount of theoretical, numerical and
experimental work of the last years, several questions are still
open. We mention here two topics that are likely to focus the
interest of researchers in the next future and are connected to
the subject of this review: dynamical heterogeneities and aging.
Simulations on high $\phi$ attractive glasses suggest that,
although the MCT correctly captures the average dynamics, the
particle motion is very heterogeneous \cite{CatesJPCM2004}, a
feature also observed at intermediate volume fractions ($\phi =
0.4$) \cite{PuertasJChemPhys2004}. New experiments are needed to
characterize dynamical heterogeneity in attractive glasses and the
new approaches described in \sref{sec:hetero} may prove useful. No
systematic experimental investigation of the aging behavior in
attractive $vs$ repulsive glasses is available yet, although
preliminary DLS data are presented in \rref{PhamPRE2004}. They
suggest a different aging in attractive and repulsive glasses: for
the former, the dynamics keep slowing down for up to 10 days,
while no significant aging effects beyond 2 days are reported for
the latter. The destabilization of the attractive glass with time,
due to activated dynamics (bond breaking), has also been
investigated in a series of numerical works
\cite{FoffiJChemPhys2004,ZaccarelliPRL2003,Saika-VoivodPRE2004}.

\section{Out-of-equilibrium soft systems}
\label{sec:outofequilibrium}

The slow dynamics of out-of-equilibrium soft materials have been
investigated experimentally for a wide variety of systems, ranging
from model systems of hard spheres to more complicated materials.
Quite generally, these experiments demonstrate that the dynamics
at very long time scales are not completely frozen: many soft
materials in the glassy phase have been shown to exhibit
ultra-slow relaxations. After reviewing these investigations, we
will describe the aging behavior that is very often associated
with the slow dynamics. Finally, internal stress relaxation will
be discussed as a possible mechanism for slow dynamics and aging
in many soft glassy materials very far from equilibrium.

\subsection{Ultra-slow relaxations}
\label{subsec:slow}

As discussed in \sref{sec:MCT}, the ideal MCT predicts the
divergence of the $\alpha$ relaxation time at the glass
transition: therefore for $\phi > \phig$ the dynamic structure
factor should exhibit a plateau after the initial decay ($\beta$
relaxation). The pioneering dynamic light scattering experiments
on colloidal hard spheres glasses of van Megen and coworkers
\cite{vanMegenPRE1998,Mortensen1999} hinted however at a second
decay of $f(q,t)$, after the plateau. Unfortunately, the
accessible time delays were too short to properly characterize
this slow relaxation. Qualitatively similar results are obtained
by Pham and collaborators \cite{PhamPRE2004}. Recent experiments
confirm indeed that a hard sphere glass is not a fully arrested
state and cage-escape plays a decisive role even in the glass
phase. Using a confocal scanning microscope, Simeonova and Kegel
\cite{SimeonovaFaradayDiscuss2003,SimeonovaPRL2004} follow in
real-time the fluorescence recovery after photo-bleaching of a
small volume within a concentrated suspensions of hard spheres and
derive from their measurements the mean squared displacement
(MSD), $<r^2>$, of the colloids. In the glass phase, they find a
subdiffusive behavior $<r^2> \sim t^{0.30}$, which extends over
seven decades in time. The largest displacements are of the order
of the particle size. This result is intriguing: it is
fundamentally different from the ideal MCT predictions and is
moreover different from the light scattering results of van Megen
\textit{et al.}, who observe an intermediate plateau of the MSD,
before detecting a further increase of $<r^2>$ at very large $t$.
Interestingly, thanks to their extremely long experiments,
Simeonova and Kegel have recently demonstrated that the
subdiffusive power law growth of the MSD corresponds actually to a
transient regime (lasting up to $\sim 10^7$ Brownian times, or
about $7$ days!), after which a plateau is eventually reached
\cite{SimeonovaPRL2004}.

Apart from these few reports, experiments probing the slow
dynamics of hard spheres systems deep in the glass phase are still
scarce. Other systems, however, provide an opportunity to
investigate in details the dynamical state of soft materials far
from the fluid-to-solid transition. Indeed, for several soft
materials, the characteristic times are intrinsically smaller than
that of hard-spheres systems, and a quench to a state far from
equilibrium may be realized more easily. These systems include
fractal colloidal gels, dilute suspension of colloidal charged
platelets, concentrated suspensions (or pastes) of colloidal
particles stabilized both by electrostatic and steric repulsions,
and compact arrangements of soft deformable particles
(multilamellar vesicles or polyelectrolyte microgels), the volume
fraction of the latter being essentially $1$. Thanks to the
multispeckle dynamic light scattering method
\cite{WongRSI1993,BartschJChemPhys1997,LucaRSI1999,ViasnoffRSI2002},
the dynamics have been probed for time delays as long as tens of
hours. A remarkable result of these experiments is that, for
several soft glassy materials, the intensity autocorrelation
function $g_2(t)-1$ relaxes completely to zero at very long $t$, a
behavior indicative of structural changes over the length scale
probed by the experiment. These findings have been obtained by
dynamic light scattering both in the multiple scattering
(diffusing wave spectroscopy, DWS \cite{DWSGeneral}) and in the
single scattering regime (DLS) \cite{Pecora}, as well as by
X-photon correlation spectroscopy (XPCS) \cite{DiatCOCIS1998}.

Diffusing wave spectroscopy measurements are performed on very
turbid samples, or on transparent systems to which probe particles
are added to increase the turbidity. Using DWS, a slow relaxation
well separated from the fast dynamics by a plateau in $g_2(t)-1$
has been measured for systems as varied as dilute suspensions of
charged clay platelets (Laponite) forming a ``Wigner glass''
\cite{KnaebelEPL2000}, colloidal pastes \cite{ViasnoffPRL2002},
and concentrated colloidal gels \cite{BissigPhysChemComm2003}. The
quantitative interpretation of the decay of $g_2(t)-1$ is not
straightforward: the standard DWS formalism allows the MSD to be
calculated from $g_2(t)$, but it requires the motion of all
scatterers to be spatially and temporally homogeneous
\cite{DWSGeneral}, a condition that is likely to be violated in
many glassy systems (see \sref{sec:hetero}). It is also worth
noting that in these experiments the length scale probed is at
most of the order of a few tens of nanometers, much smaller than
the typical size of the sample constituents.

Qualitatively similar features have been observed in single
scattering experiments on a variety of systems, including fractal
colloidal gels with strong and irreversible
\cite{CipellettiPRL2000} or weak, thermoreversible
\cite{SolomonPRE2001} bonds between the particles, compact
arrangements of soft elastic spheres \cite{RamosPRL2001} or of
emulsion droplets \cite{CipellettiFaradayDiscuss2003},
polycrystals made of block copolymer micelles
\cite{CipellettiFaradayDiscuss2003}, and suspensions of Laponite
\cite{BellourPRE2003,AbouPRE2001}. Note that DLS probes length
scales much larger than those explored in a DWS experiment,
typically up to the size of the sample constituents or more. The
full decay of the dynamic structure factor extracted from DLS
experiments is therefore particulary significative, since it is
indicative of slow rearrangements that eventually change the
system configuration on length scales comparable to or larger than
the size of its constituents. An additional valuable feature of
DLS is the possibility to easily measure how the dynamics depend
on length scale. This is accomplished by changing the $q$ vector
at which the experiment is performed. Several different
$q$-dependencies of the characteristic time $\tau$ for the slow
relaxation have been measured. Abou and coworkers
\cite{AbouPRE2001} have found $\tau \sim q^{-2}$ for a suspension
of clay particles (Laponite), as expected for a diffusive process,
such as cage-escape mechanisms. However, less intuitive scaling
laws have also obtained by several groups on various samples.
Solomon and coworkers \cite{SolomonPRE2001} reported $\tau \sim
q^{-0.5}$ for a colloidal gel of adhesive particles, while Bellour
and coworkers have found ---for the same material as Abou
\textit{et al.} but in a different aging regime, see also
subsection \ref{subsec:aging}--- $\tau \sim q^{-1.3}$. These
results remain largely unexplained.

\begin{figure}
\begin{center}
\includegraphics[scale=0.9]{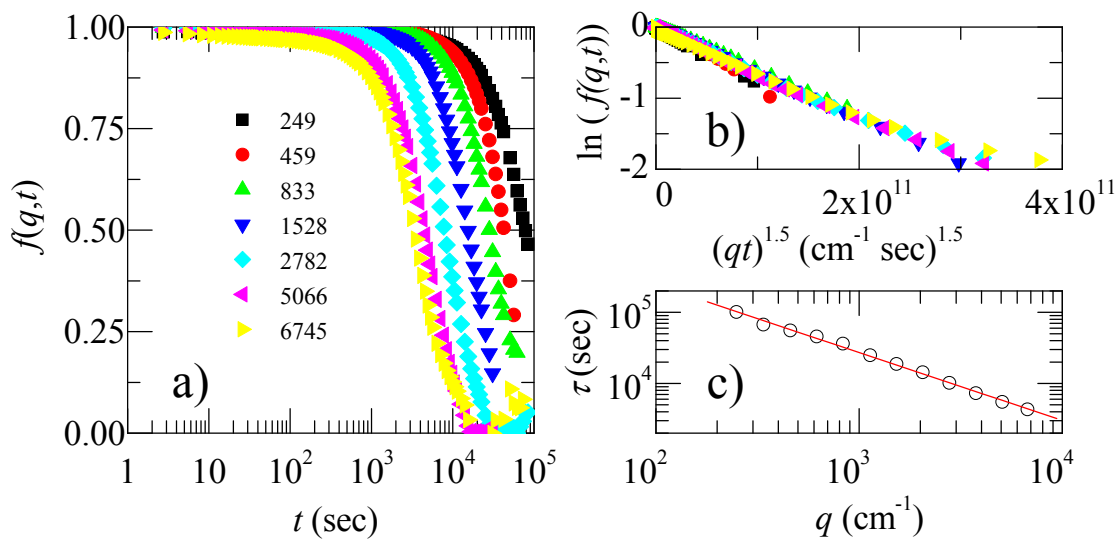}
\end{center}
\caption{\label{fig:compexp} a): dynamic structure factor measured
simultaneously at various scattering vectors for a fractal gel
made of strongly attractive polystyrene particles, with $\phi = 5
\times 10^{-4}$. The curves are labelled by the magnitude of the
scattering vector in $\textrm{cm}^{-1}$. b): the sama data as in
a) are plotted as $\ln[f(q,t)]$ $vs$ $(qt)^{1.5}$. All data
collapse onto a straight line, thus demonstrating that $ f(q,t)
\sim \exp[-(t/\tau)^p]$, with $p=1.5$. c): double logarithmic plot
of the $q$ dependence of the relaxation time $\tau$. The line has
a slope of $-0.96 \pm 0.02$, suggesting a $\tau \sim q^{-1}$
scaling. To avoid overcrowding the figure, not all the curves that
were fitted to obtain the data shown in c) are plotted in a) and
b). Data taken from reference \cite{CipellettiPRL2000}. }
\end{figure}

An intriguing $\tau \sim q^{-1}$ scaling has been measured for a
wide variety of materials, including both attractive systems
(fractal colloidal gels \cite{CipellettiPRL2000}) and repulsive
systems (compact arrangements of soft elastic spheres
\cite{RamosPRL2001} or of emulsion droplets
\cite{CipellettiFaradayDiscuss2003}, a polycrystal made of
copolymer micelles \cite{CipellettiFaradayDiscuss2003}, and
Laponite in XPCS experiments that probe the high $q$ regime
\cite{BandyopadhyayPRL2004}). This scaling suggests that the slow
dynamics be due to a ``ballistic'' motion of the particles, in the
sense that the average particle displacement grows linearly with
time (this ultra-slow regime should not be confused with the
ballistic motion of colloids observed on very short time scales,
before the onset of Brownian motion \cite{WeitzPRL1989}). A very
unusual shape of the dynamic structure factor accompanies always
this scaling: $f(q,t)$ decays as a ``compressed exponential'',
i.e. $ f(q,t) \sim \exp[-(t/\tau)^p]$, with a ``compressing''
exponent larger than one, $p\approx 1.5$. \Fref{fig:compexp}
illustrates both the $\tau \sim q^{-1}$ scaling and the compressed
exponential shape of $ f(q,t)$ for the colloidal fractal gel
studied in \cite{CipellettiPRL2000}. Note that the same compressed
exponential shape was observed in reference \cite{BellourPRE2003},
although with a somehow smaller value of $p\approx 1.35$.

Because the relaxation is faster than exponential and due to the
$\tau \sim q^{-1}$ scaling, all models based on cage-escape
processes must be ruled out. Instead, we have proposed
\cite{RamosPRL2001,CipellettiPRL2000,CipellettiFaradayDiscuss2003}
the dynamics to be due to randomly distributed internal stress
sources acting on the sample, whose response is that of an elastic
solid. Simple scaling arguments lead then to the observed peculiar
shape of the dynamic structure factor
\cite{CipellettiFaradayDiscuss2003}. A more refined model based on
the microscopic description of local rearrangement events has been
developed for the case of a colloidal gel \cite{BouchaudEPJE2001}.
Experiments testing the notion of internal stress as the driving
force for the slow dynamics are reviewed in
sub\sref{subsec:internalstress}. Finally, we note that there are
strong indications that in this type of systems the dynamics is
heterogeneous. The unusual dynamics reported above may therefore
be due to a series of discrete rearrangement events, as it will be
discussed in more details in \sref{sec:hetero}.

\subsection{Aging}
\label{subsec:aging}

The properties of out-of-equilibrium systems continuously change
with time as the system slowly evolves towards its equilibrium
configuration. As a consequence, correlation functions, e.g.
measured by dynamic light scattering, and response functions, e.g.
stress or strain relaxation in rheological experiments, depend not
only on time delay, $t$, as for time-translation invariant
systems, but also on the waiting time or sample age, $\tw$.
Typically, the dynamics becomes progressively slower as the sample
ages and the system spends more and more time in the metastable
states it visits, the energy of such states becoming increasingly
lower (see references \cite{Bouchaud2000,VincentLectNotesPhys1997}
for an introduction to aging phenomena). Very generally, age is
defined as the time elapsed since the sample was quenched from a
fluid to an out-of-equilibrium solid state. For hard condensed
matter systems (e.g. structural, polymeric or spin glasses), this
is usually realized by means of a temperature quench. By contrast,
soft materials are very often (albeit not always) initialized by
applying a strong mechanical shear; hence age is defined as the
time interval between the end of the applied shear and the
beginning of the measurement. Aggregation phenomena leading to a
percolated network may also provide a route to an
out-of-equilibrium state: in this case, $\tw = 0$ is taken as the
percolation (or gelation) time.

The first experimental reports of aging phenomena in soft
condensed matter are light scattering experiments performed on
hard spheres colloidal glass \cite{vanMegenPRE1998,Mortensen1999}.
Although the accessible time window in these experiments was too
short to investigate in details the aging dynamics, the data
unambiguously show a slowing down of the dynamics as the sample
ages, when the volume fraction of the sample is larger than
$\phig$. By contrast, no dependence of the dynamic structure
factor with sample age can be detected when $\phi$ is smaller than
$\phig$. Hence, the results of van Megen and coworkers nicely
demonstrate the dramatic dynamical differences between a sample in
the supercooled fluid state and one in the glass state, in spite
of the minute differences in sample composition. More recently,
Simeonova and Kegel \cite{SimeonovaPRL2004}, using the fluorescent
recovery after photobleaching technique, have found clear evidence
of aging behavior in hard sphere glasses, since they measure a
particle MSD whose growth slows down with increasing $\tw$.
Similar results have been obtained by Courtland and Weeks
\cite{CourtlandJPCM2003}, who used confocal microscopy to follow
in real-time the three-dimensional motion of individual particles
in a colloidal glass. They find that the MSD initially grows
linearly with time, then tends to plateau, and finally exhibits a
slight upturn at very large time (similarly to what observed for
the cage escape process in supercooled fluids). For glasses, the
time scale of the upturn increases with $\tw$. They also find that
the dynamics is both spatially and temporally heterogeneous,
although characterizing dynamical heterogeneities is more
difficult than in supercooled samples, because the particle motion
is very restrained and the statistics is poorer (more details on
dynamical heterogeneities can be found in \sref{sec:hetero}). At
all ages the most mobile particles form clusters; however no
obvious evolution of the cluster size or morphology with age is
observed. To conclude the overview of experiments on the aging of
hard spheres systems, we recall the preliminary dynamic light
scattering results in \rref{PhamPRE2004}, which compare the aging
dynamics of a repulsive hard sphere colloidal glass to those of an
attractive glass (see \sref{sec:MCT}) and suggest different
behaviors: the repulsive glass seems to stop aging after about one
day, while the dynamics of the attractive glasses keep on slowing
down after $10$ days.

Aging behavior has been most extensively investigated in soft
materials which greatly depart from a hard spheres model system.
These materials include colloidal pastes made of concentrated
silica \cite{DerecCRAS2000,DerecPRE2003} or polystyrene particles
\cite{ViasnoffPRL2002} (both stabilized by a combination of
electrostatic and steric repulsions), diluted
\cite{CipellettiPRL2000} or more concentrated
\cite{BissigPhysChemComm2003} fractal colloidal gels of
polystyrene beads, diluted suspensions of charged clay platelets
(Laponite) \cite{BellourPRE2003,AbouPRE2001} (to which polystyrene
beads may be added to perform DWS experiments
\cite{KnaebelEPL2000}), compact arrangements ($\phi \sim 1$) of
surfactant multilamellar vesicles \cite{RamosPRL2001},
polyelectrolyte microgel particles \cite{CloitrePRL2000}, or
polycrystals of copolymeric micelles
\cite{CipellettiFaradayDiscuss2003}.

For these systems, rheological tests are very often employed to
probe aging phenomena, in addition to light scattering
experiments. Rheological experiments, e.g. measurements of the
stress relaxation following a step strain in the linear regime,
are often compared to available theories, one of the most
successful being the Soft Glassy Rheology (SGR) model of Sollich
and coworkers \cite{SollichPRL1997,SollichPRE1998}. This model,
derived from the trap model of Bouchaud
\cite{Bouchaud2000,MonthusJphysA1996} accounts for the rheology of
out-of-equilibrium materials. In brief, in the trap model
non-interacting particles evolve through a hopping mechanism in an
energy landscape with wells of depth $E$. The distribution of well
depths is fixed and the evolution of $P(E,t)$, the probability for
a particle to be in a trap of depth $E$ at time $t$, is governed
by thermally activated hopping. This model leads to aging
phenomena when the average trapping time diverges. In the SGR
model, the material is divided into elements which yield above a
critical yield strain $l_y$, thus relaxing stress. Yield events
(i.e. rearrangements of the particles) are seen as hops out of the
trap, and yield energy is identified as the trap depth $E$. The
activation barrier is $E=E_y-\frac{1}{2}k l^2$ where
$E_y=\frac{1}{2}k l_y^2$ is the maximum elastic energy before
yielding, $l$ is the local shear strain of an element, and $k$ is
an elastic constant. The link between the microscopic parameters
and the macroscopic strain applied to the sample is
straightforward: in between rearrangements, the local strain
follows the macroscopically imposed strain $\gamma$. In
\rref{FieldingJRheol2000}, Fielding, Sollich and Cates study the
role of aging in the rheology of soft glassy materials. They
describe several rheological tests suitable to investigate the
(linear and non-linear) rheology of aging soft materials, whose
features are qualitatively similar to experimental results.

Experimentally, aging behavior has been observed in DWS
\cite{KnaebelEPL2000,ViasnoffPRL2002,BissigPhysChemComm2003},
single light scattering
\cite{RamosPRL2001,CipellettiPRL2000,BellourPRE2003,AbouPRE2001},
and linear rheology
\cite{RamosPRL2001,DerecCRAS2000,DerecPRE2003,CloitrePRL2000}
experiments. In all cases, the characteristic time $\tau$ of the
correlation functions (for light scattering experiments), of the
stress relaxation (in a step strain experiment), or of the strain
evolution (in a creep experiment) increases continuously with
sample age. However, very different aging laws $\tau(\tw)$ have
been found for the various materials investigated. In many cases a
power law $\tau \sim \tw^{\mu}$ is observed, but the aging
exponent $\mu$ may significantly vary. Indeed, values of about $1$
\cite{KnaebelEPL2000,ViasnoffPRL2002,CipellettiPRL2000,CloitrePRL2000},
smaller than $1$ ($0.5 < \mu <1$)
\cite{RamosPRL2001,DerecCRAS2000,DerecPRE2003}, and larger than
$1$ ($\mu \simeq 1.4$ and $\mu \simeq 1.8$ in references
\cite{BissigPhysChemComm2003} and \cite{BandyopadhyayPRL2004},
respectively) have all been observed, reflecting ---in the
language of glasses--- full aging, sub-aging and hyper-aging,
respectively. A peculiar, very fast aging regime has been observed
both in fractal colloidal gels \cite{CipellettiPRL2000} and in
Laponite samples \cite{BellourPRE2003,AbouPRE2001}, for which
$\tau$ is found to grow exponentially with sample age. This fast
aging regime is eventually followed by a slower growth of $\tau$,
corresponding to full aging. Note that the work by Bellour {\it et
al.} \cite{BellourPRE2003}, who studied the aging of Laponite over
a wide interval of $\tw$ and observed both regimes, rationalizes
the apparently conflicting findings of references
\cite{KnaebelEPL2000} and \cite{AbouPRE2001}, which were probing
only the exponential and the full aging regime, respectively.

The experimental work reviewed in this section clearly shows that,
albeit aging behavior is a ubiquitous feature of
out-of-equilibrium soft materials, the detailed evolution of the
dynamics with $\tw$ greatly depends on the particular system that
is investigated. The large spectrum of aging behaviors measured
experimentally hints at differences in the microscopic mechanisms
at play in the various samples and stands as a challenge for more
detailed theories.

\subsection{Internal stress relaxation}
\label{subsec:internalstress}

In subsec.~\ref{subsec:slow}, we have mentioned that a large
variety of soft disordered materials were found to exhibit an
unusual slow dynamics, with  a compressed exponential shape for
the dynamic structure factor ($f(q,t) \sim \exp[-(t/\tau)^p]$,
with $p\approx1.5$ and $\tau \sim q^{-1}$), indicative of
ballistic motion
\cite{RamosPRL2001,CipellettiPRL2000,CipellettiFaradayDiscuss2003,BandyopadhyayPRL2004}.
We have argued that these peculiar features could be rationalized
with simple arguments, based on the concept that the dynamics be
due to the relaxation of internal stresses. Several experimental
observations hint at the key role played by internal stresses in
leading to the final relaxation of $f(q,t)$. Cipelletti $et$ $al.$
and Manley $et$ $al.$ \cite{CipellettiPRL2000,ManleySilicaGel2004}
have observed particle syneresis in colloidal gels made of
attractive particles. When the gel is strongly anchored to the
container walls (as it is usually the case), tensile stress builds
up in the sample as a result of the decrease of the interparticle
distance, due to syneresis. Further evidence for the role of
internal stress is provided by experiments on concentrated
emulsions, whose dynamics was initialized by centrifugation
\cite{CipellettiFaradayDiscuss2003}. The dynamics is
systematically faster in the direction of the centrifugation
acceleration, along which most of the internal stress has been
presumably built in. For Laponite suspensions, Bandyopadhyay and
coworkers argue that internal stresses are generated by the
increase of the interparticle repulsion, due to the dissociation
of ions at the surface of the particle, as revealed by the
increase of the conductivity of the suspension with sample age
\cite{BandyopadhyayPRL2004}. Finally, similar dynamics were
observed for a micellar polycrystal
\cite{CipellettiFaradayDiscuss2003} and compact arrangements of
multilamellar vesicles (MLVs) \cite{RamosPRL2001}, two systems for
which the fluid-to-solid transition is induced by a temperature
jump. For these samples, we expect internal stress to be built up
due to the rapid growth of randomly oriented crystallites or MLVs,
respectively. Interestingly, linear rheology measurements on the
MLVs \cite{RamosPRL2001} show that the characteristic time of the
mechanical response of this material follows exactly the same
aging law as the characteristic time of the final decay of
$f(q,t)$. Since rheology probes the response to an external
stress, while the final relaxation of $f(q,t)$ is ascribed to
internal stress, this concordance gives further support to the
concept of stress relaxation as a key ingredient in determining
the dynamics of disordered systems and their evolution with sample
age.

Dense arrangements of MLVs \cite{RamosPRL2001,RamosEPL2004}
provide ideal samples to test in great details the notion of
internal stress as a driving force for the slow dynamics. As
mentioned above, the fluid-to-solid transition of this material
can be driven by a temperature jump. This has to be contrasted
with most soft glassy systems, where the fluid-to-solid transition
is obtained upon cessation of a large applied shear, which
certainly influences the initial configuration of internal stress.
For the MLVs, the physical origin of the internal stress is the
elastic energy stored in the deformation of the vesicles, which
are expected to depart from a spherical, elastically relaxed
shape, due to their rapid and disordered growth at the
fluid-to-solid transition. On the other hand, the linear elastic
modulus $G_0$ is equal to the density of elastic energy stored by
the material when it is deformed in the linear regime
\cite{RamosEPL2004}. Therefore, both the internal stress and $G_0$
share the same microscopic origin, i.e. the deformation of the
MLVs. It follows that the slow dynamics should depend crucially on
$G_0$, if indeed they are driven by the relaxation of internal
stress. In order to test this conjecture, we have recently
measured the aging dynamics of MLVs samples for which $G_0$ was
varied over more than one order of magnitude
\cite{RamosUnpublished2004}. We find that the slow dynamics is
faster for systems with a higher elastic modulus, in agreement
with the hypothesis that the higher $G_0$ the larger the internal
stress. Remarkably, the slow dynamics and the aging can be
entirely described by the evolution of an effective viscosity,
$\eta_{eff}$, defined as the characteristic time measured in a
stress relaxation rheology test times $G_0$. The concept of
effective viscosity is found to be robust, since at all time
$\eta_{eff}$ is independent of $G_0$, of elastic perturbations,
and of the rate at which the sample is driven from the fluid to
the solid state. A simple model that links $\eta_{eff}$ to the
internal stress created at the fluid-to-solid transition is
proposed. In this model, the ballistic motion of the MLVs results
from a balance between a driving force, associated to the local
internal stress acting on a region containing several MLVs, and a
viscous drag. (Note that this model is similar to the one proposed
in reference \cite{CloitrePRL2003} to describe the fast dynamics
associated with the rattling of soft microgel particles within the
cage formed by their neighbors.) In this picture, aging results
from a weakening of the driving force, due to the progressive
relaxation of internal stress. Indeed, we find that $G_0$ slowly
decreases with $\tw$, as expected if the measured elastic modulus
is the sum of a (constant) ``equilibrium'' elastic modulus
(corresponding to an ideal, totally relaxed configuration of the
sample, where all MLVs have a spherical shape) and the internal
stress, which decreases with $\tw$.

Similar arguments should apply also to other soft systems, such as
concentrated emulsions, whose elasticity, similarly to that of the
MLVs samples, depends on the deformation of their constituents. By
contrast, a very different behavior is observed for hard
particles, for which the elasticity results essentially from
excluded volume interactions (entropic origin). Derec \textit{et
al.} study the age-dependent rheology of a colloidal paste made of
silica particles \cite{DerecPRE2003}. They find that the elastic
modulus of the paste increases logarithmically with sample age,
defined as the time elapsed since the system has been fluidified
by a strong mechanical shear, while the elastic modulus of a MLV
sample decreases. The authors suggest a link between the aging
that they observe in stress relaxation experiments and the
spontaneous increase of the elastic modulus. A phenomenological
model which incorporates as main ingredients spontaneous aging and
mechanical rejuvenation
\cite{DerecCRAS2000,DerecPRE2003,DerecEPJE2001} is able to
reproduced the essential features of the various rheological tests
performed experimentally, in particular the stress relaxation
experiments, although in the model the elastic modulus of the
sample is fixed. The same phenomenological model gives also
results in good agreement with start-up flow experiments
\cite{DerecPRE2003}, where the time evolution of the stress is
measured as a constant shear rate is imposed to the sample. They
find both experimentally and theoretically that the stress
overshoots before the system actually starts to flow. The
amplitude of the overshoot increases with sample age, thus
suggesting that the older the sample the larger the stress that
has to be stored before flow can occur. This suggests that the
system strengthens with age, in contrast with the decrease of
$G_0$ observed for the MLVs.

Interestingly, a similar link between slow dynamics, aging
behavior and stress relaxation has been proposed to explain the
dynamics of a gently vibrated granular pile \cite{KablaPRL2004}.
Kabla and Debr\'{e}geas measure by multispeckle DWS the two-time
intensity autocorrelation function of the light scattered by the
pile, $g_2(t,\tw)$. The sample is vibrated by applying ``taps'',
whose amplitude is small enough not to induce macroscopic
compaction, but large enough to trigger irreversible particle
displacements on a microscopic scale. The characteristic decay
time of $g_2(t,\tw)$ increases roughly linearly with $\tw$, in
surprising analogy with many glassy systems. By invoking arguments
similar to those of Bouchaud's trap model
\cite{Bouchaud2000,MonthusJphysA1996}, the authors explain this
slowing down as the result of a slow evolution of the
(gravitationally-induced) stress distribution in the pile. Weak
grain contacts are progressively replaced by stronger contacts,
leading to the observed aging of the dynamics and the
strengthening of the pile.

The literature reviewed in this section indicates that internal
stress and its time evolution are quite general ingredients in
attempts to explain the slow dynamics and the aging of many glassy
systems. However, internal stress may play very different roles,
either acting as a driving force for the dynamics (as proposed,
e.g., for the MLVs), or evolving in response to other processes
(e.g. thermal activation or applied vibrations). The experiments
reviewed here show that the role of internal stress, as well as
the evolution of the elastic response during aging, may depend
also on the way the samples are initialized and on the microscopic
origin of the elasticity. Indeed, the internal stress can not be
the same when a system is fluidized by applying a mechanical shear
or, on the contrary, without perturbing it mechanically.
Similarly, different aging behaviors of the stress distribution
and the elasticity are to be expected in samples where the
elasticity is due to the deformation or the bending of the
individual constituents, or where it derives from excluded volume
interactions.

\section{Response to an external perturbation}
\label{sec:perturb}

A great amount of experimental, numerical and theoretical work has
been devoted to investigating the response of glassy soft
materials to an external perturbation, usually a mechanical one.
There are several reasons justifying such a broad interest. On one
hand, these materials are ubiquitous in industrial applications,
where their mechanical properties are of primary importance. On
the other hand, applying a (large) mechanical perturbation is a
way to modify the dynamical state and the aging behavior of a soft
glass: in this case the dynamics may depend on the mechanical
history, much as the dynamics of a molecular or spin glass depends
on its thermal history. This analogy lies on the similar role of
temperature and, e.g., strain in driving the fluid-to-solid
transition, as proposed by Liu and Nagel \cite{LiuNature1998}.
Finally, there is an intense debate on the relationship between
response functions (in the linear regime) and correlation
functions in glassy systems and the break down of the Fluctuation
Dissipation Theorem (FDT).

In this section, we will first review experiments probing the
response of a system to a strong perturbation (non-linear regime),
and then discuss experimental work addressing the possible
violation of the FDT.

\subsection{Non-linear regime: rejuvenation and overaging}
\label{subsec:nonlinear}

Habdas \textit{et al.} have performed original experiments
\cite{HabdasEPL2004} where they measure the average velocity and
the velocity fluctuations of a magnetic bead submitted to a force,
$f$, and immersed in a dense colloidal suspension (in the
supercooled fluid state). As also observed in simulations
\cite{HastingsPRL2003}, they find that the average velocity varies
as a power law with $f$. They also find a threshold force $f_0$
for motion to be observed, which initially increases with the
colloid volume fraction, but eventually saturates, suggesting that
$f_0$ does not diverge at the glass transition. Moreover, they do
not observe an increase of velocity fluctuations when approaching
the glass transition, as one might have expected if the probe
environment became more heterogeneous. They suggest that $f_0$ be
related to the strength of the cage and that the existence of
$f_0$ hints at local jamming even if the colloidal suspension is
globally in a liquid phase. Two interesting extensions of this
work would be \textit{i)} to perform the same type of measurements
in the glass phase and \textit{ii)} to look at the rearrangements
of the particles due to the forced motion of the magnetic probe.
We note that experiments similar to \textit{ii)} have been
conducted in a two-dimensional granular material where one follows
the motion of all grains in response to the forced motion of one
grain \cite{KolbPRE2004}. Note that in the experiment by Habdas
\textit{et al.} the motion of the magnetic probe is much larger
than the Brownian motion of the surrounding particles. As a
consequence, this experiment probes
---at a microscopic level--- the non-linear rheology of the
suspensions .

Macroscopic non-linear rheology experiments have been performed on
a variety of systems. Although it has been known for several
years, and has been widely used in experiments, that the dynamics
of soft glassy materials can be initialized by submitting them to
a strong shear, recent experiments have investigated in more
details the effects of shear on the slow dynamics and the aging of
these materials. One important issue is to understand how the slow
and age-dependent evolution of the rheological properties
correlates with the mechanical history of the materials. A first
answer is provided by Cloitre \textit{et al.}
\cite{CloitrePRL2000}, who have studied a paste made of soft
microgel colloidal particles. They measure the strain recovery and
the creep of a sample (initially fluidified by applying a stress
larger than the yield stress, $\sigma_y$) submitted to a ``probe''
stress $\sigma$ of variable amplitude. They find that the time
evolution of the strain is age-dependent but that all the
experimental data collapse on a master curve when the time is
normalized by $\tw^{\mu}$. The aging exponent $\mu$ decreases
continuously from $1$ (full aging) to $0$ (no aging) as $\sigma$
increases ($\mu=0$ for $\sigma=\sigma_y$). They propose that $\mu$
can be used to quantify the partial mechanically-induced
rejuvenation of the sample.

Light scattering has also been used extensively to investigate the
interplay between slow dynamics and shear rejuvenation. Using the
recently introduced light scattering echo technique
\cite{HebraudPRL1997,HohlerPRL1997}, Petekidis and coworkers
\cite{PetekidisPRE2002} have looked at the effect of an
oscillatory shear strain on the slow relaxation of a colloidal
glass of hard spheres, finding support for a speedup of the slow
dynamics due to shear. Bonn and coworkers \cite{BonnPRL2002} have
measured by DWS the characteristic relaxation time $\tau$ of a
glassy suspension of Laponite particles, to which a shear of rate
$\dot{\gamma}$ has been applied. Similarly to the results of
\rref{PetekidisPRE2002}, they find that $\tau$ decreases after
imposing a shear; moreover $\tau$ is smaller for larger
$\dot{\gamma}$. The observed shear rejuvenation is in agreement
with theoretical predictions \cite{BerthierPRE2000}.

More quantitatively, Ozon \textit{et al.} \cite{OzonPRE2003} have
shown using clay particles (smectite) that partial rejuvenation
depends only on the amplitude of the applied strain and not on its
frequency nor its duration. They find that the relative decrease
of the relaxation time due to a sinusoidal shear varies
exponentially with strain amplitude, $\gamma$, implying that the
mechanical energy input is proportional to $\gamma$. This result
is in contradiction with the Soft Glassy Rheology model of Sollich
\textit{et al.} \cite{SollichPRL1997,SollichPRE1998}, which
assumes a $\gamma^2$ dependence, resulting from a local elastic
response.

\begin{figure}
\begin{center}
\includegraphics [scale=0.9]{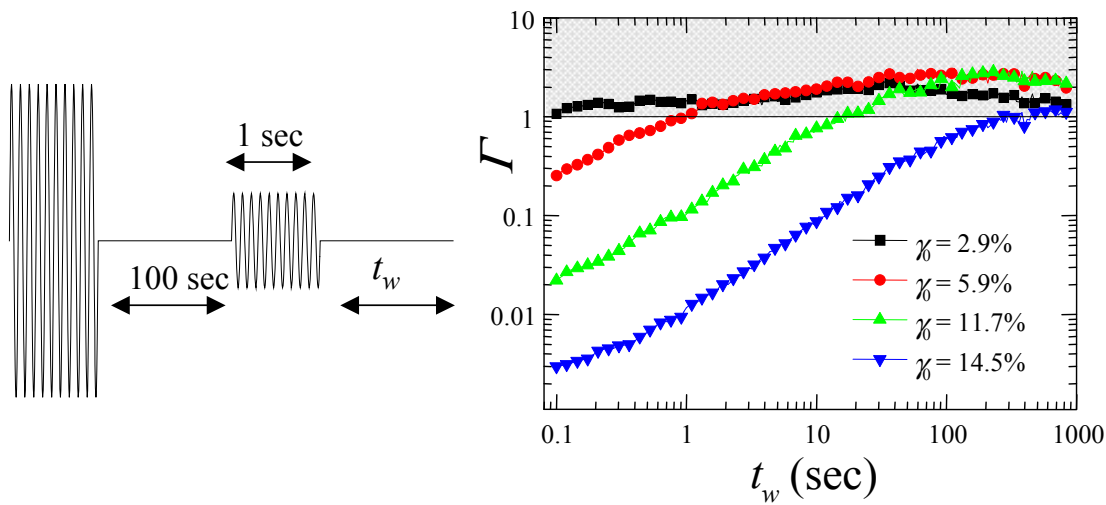}
\end{center}
\caption{\label{fig:overaging} (Left) Strain history and (Right)
normalized relaxation time $\Gamma$ (ratio of the relaxation time
of the perturbed sample over relaxation time of the unperturbed
sample) as a function of sample age. Rejuvenation corresponds to
$\Gamma < 1$ and overaging corresponds to $\Gamma
> 1$ (gray region). Curves are labelled by strain amplitudes. Overaging occurs
for low strain amplitude. (Adapted from reference
\cite{ViasnoffFaradayDiscuss2003})}
\end{figure}

This discrepancy is only but an example of how subtle and
counterintuitive the mechanisms for shear rejuvenation may be. By
investigating in detail the dynamics of a colloidal paste
subjected to an oscillatory mechanical strain of variable
duration, amplitude and frequency, Viasnoff and Lequeux
\cite{ViasnoffPRL2002,ViasnoffFaradayDiscuss2003} have indeed
demonstrated that the characteristic relaxation time of the
mechanically perturbed sample (as measured by DWS)  may be either
smaller or larger than that of an unperturbed sample. The former
case corresponds to rejuvenation, while the latter has been termed
``overaging''. An illustration of the rejuvenation and overaging
effects is shown in figure \ref{fig:overaging}. Note that
overaging is reminiscent to observations of the dynamics of a foam
by Cohen-Addad and H\"ohler \cite{CohenAddadPRL2001}, who have
found that the bubble dynamics is strongly slowed down after
applying a transient shear. Viasnoff and Lequeux interpret their
findings using Bouchaud's trap model for glassy dynamics
\cite{Bouchaud2000,MonthusJphysA1996} (see
subsec.~\ref{subsec:aging} for a short description), assuming that
shear is strictly equivalent to temperature. They calculate the
probability for a particle to be in a trap of depth $E$ a time $t$
after a temperature jump of small amplitude. They find that the
$T$-jump (correspondent to the shear perturbation in their
experiments) modifies the lifetime distribution : at large times,
the deep traps  may be overpopulated compared to the unperturbed
case, while populations of the shallow traps are equal in both
cases. This change of the distribution of life times for the traps
leads to an average relaxation time which is longer for the
perturbed sample, and hence to overaging. Physically, overaging
demonstrates that a moderate shear could help the system to find a
more stable configuration, by allowing it to explore more rapidly
a larger portion of the energy landscape.

Shear rejuvenation may lead to very peculiar rheological
responses. Coussot, Bonn and coworkers have indeed observed for
several systems \cite{CoussotPRL2002,BonnEPL2002} a critical
stress above which the viscosity $\eta$ continuously decreases
with time and below which $\eta$ increases until flow is totally
arrested. The authors interpret the viscosity bifurcation as
resulting from a competition between aging and rejuvenation and
propose a simple phenomenological model that captures the main
experimental observations.

An alternative way to perturb a system have been recently studied
by Narita \textit{et al.} \cite{NaritaEPJE2004}. The authors have
measured by DWS the slow dynamics of concentrated colloidal
suspensions upon drying. They consider that in this experimental
configuration a uniaxial compressive stress is generated in the
sample. They attempt to map their previous findings of
mechanically-induced rejuvenation and overaging on their
experimental results. Implicitly, they assume that the effects on
the internal stress in the aging dynamics of the sample are
similar to those of an external strain. Detailed experiments are
clearly needed to address this issue.

Finally, recent experiments of Simeonova and Kegel
\cite{SimeonovaPRL2004} demonstrate that soft glasses can be
considerably perturbed also by gravitational stress. By changing
the solvent in which the colloids are suspended, the density
difference between the colloids and the solvent, $\Delta \rho$, is
varied. Two systems with gravitational lengths $h$ equal
respectively to $100$ and $10$  $\mu$m
($h=k_{\rm{B}}T/(\frac{4\pi}{3}R^3 \Delta \rho)$, with $k_{\rm B}$
the Boltzmann constant and $R$ the radius of the colloids) are
used. They find that gravity accelerates significantly the aging
of the colloidal glass. Thus, a direct parallel with the overaging
effect can be drawn: gravity, similarly to a moderate shear, can
help the system to find a more stable configuration. On the other
hand, gravity prevents a full exploration of the configuration
space, ultimately suppressing crystallization. Indeed, space
experiments have shown that hard sphere suspensions that are in a
glass phase on earth do crystallize in microgravity conditions
\cite{ZhuNature1997}. How gravity couples to the particle
rearrangements on a microscopic scale remains an open issue.

\subsection{Violation of the Fluctuation Dissipation Theorem and effective temperature}
\label{subsec:FDT}

For systems at equilibrium, the response to an external
perturbation is related to the correlation function of the
observable to which the perturbation field is conjugated. Let us
consider an observable $A$ and its normalized time autocorrelation
function $C(t)= \frac{\langle A(t_0+t) A(t_0) \rangle}{\langle
A(t_0) A(t_0) \rangle}$. If the system is perturbed by a field $h$
conjugated to $A$, the response function is $R(t)= \frac{\delta
A(t)}{\delta h(t_0)}$. At equilibrium, the fluctuation-dissipation
theorem (FDT) relates the time derivative of the correlation
function with the response: $\frac{dC}{dt}=-k_{\rm B} T R(t)$
where $k_{\rm B}$ is the Boltzmann constant and $T$ is the system
temperature. Two common examples of the FDT are the Nyquist
formula that relates the voltage noise across a resistor to the
electrical resistance and the Stokes-Einstein relation, which
relates the diffusion of a particle in a solvent to the solvent
viscosity. In deriving the FDT, time-translation invariance is
required: as this assumption is not fulfilled for
out-of-equilibrium systems, for them the FDT does not hold. For
out-of-equilibrium systems, it has been proposed
\cite{CugliandoloPRL1993,CugliandoloPRE1997} that the FDT could be
generalized by introducing an effective temperature, $T_{\rm
eff}$. The time derivative of the correlation function and the
response function are then related through $\frac{dC}{dt}=-k_{\rm
B} T_{\rm eff} R(t)$. Note that $T_{\rm eff}$ is expected to
depend on the observation time scale and on the sample age and to
be higher than the temperature $T$ of a bath with which the
out-of-equilibrium material is in thermal contact.

In the past, violations of the FDT have been experimentally
observed for structural glasses \cite{GrigeraPRL1999}, spin
glasses \cite{HerissonPRL2002}, and polymer glasses
\cite{BuissonEPL2003}; moreover, an effective temperature has been
introduced to describe granular materials
\cite{D'AnnaNature2003,OjhaNature2004}. By contrast, experiments
testing the FDT in out-of-equilibrium soft materials are still
very rare. Using available data from the literature, Bonn and
Kegel \cite{BonnJChemPhys2003} examine the generalized
Stokes-Einstein relation, $D(\omega)=k_B T / [6 \pi \eta(\omega)
R]$, for hard sphere suspensions. Here $D(\omega)$ and $R$ are the
frequency-dependent diffusion coefficient and the radius of the
particles, respectively, and $\eta(\omega)$ is the
frequency-dependent viscosity of the suspension. They derive
$D(\omega)$ from dynamic light scattering experiments and
$\eta(\omega)$ from rheology measurements of either the high shear
rate viscosity of the suspension, when considering the short-time
diffusion coefficient, or the complex modulus in the linear
regime, when considering the long-time diffusion coefficient. They
find that, in a wide range of volume fractions (between $0.032$
and $0.59$), the FDT is obeyed ($T_{\rm eff}=T$) for the short
time diffusion coefficient, i.e., at the larger volume fractions,
for the particle dynamics within the cage. By contrast, for
supercooled suspensions the FDT is strongly violated at longer
times. The departure from FDT is quantified by $\frac{T_{\rm
eff}}{T}=\widetilde{\eta}(\omega) \widetilde{D}(\omega)$, where
$\widetilde{D}$ and $\widetilde{\eta}$ are reduced parameters
given in units of the diffusion coefficient of the particle at
infinite dilution and of the solvent viscosity, respectively. They
deduce that $\frac{T_{\rm eff}}{T}$ can be as large as $70$ at low
frequencies and for a volume fraction $\phi=0.52$; $\frac{T_{\rm
eff}}{T}$ is found to decrease from $\sim 70$ to  $\sim 10$ as
$\phi$ increases from $0.52$ to $0.56$. As expected theoretically,
they find that $T=T_{\rm eff}$ at high frequency and that at low
frequency $T_{\rm eff}$ is larger than $T$ and decreases
continuously with $\omega$ following a power law. Finally, we
point out that, although the violation of the FDT is clearly
evidenced in this work, the exact values of the effective
temperatures should be taken with extreme caution since they are
derived from experiments carried out with different particles.

Abou and Gallet \cite{AbouPRL2004} have measured the effective
temperature of a colloidal glass of clay particles (Laponite),
using a modified Stokes-Einstein relation (the detailed
theoretical formalism for the diffusion of a particle in an aging
medium can be found in
\rsref{PottierPhysicaA2003,PottierPhysicaA2004}). They seed the
colloidal glass with micron-sized particle probes that can be
tracked using light microscopy, and use these probes as a
thermometer by measuring both the mean squared displacement and
the mobility of the beads. From these measurements, they extract
an effective temperature $T_{\rm eff}$ at a fixed frequency of $1$
Hz. They find a non-monotonic variation of $T_{\rm eff}$ with
sample age: $T_{\rm eff}$ is equal to the bath temperature $T$ for
a young sample, it increases up to $\simeq 1.8 T$ at intermediate
age ($\tw \simeq 150$ min) and then decreases when the sample
further ages. The authors relate the non-monotonic variation of
$T_{\rm eff}$ with sample age to the evolution of the
characteristic time for the relaxation of the colloidal glass as
measured by dynamic light scattering, and argue that $T>T_{\rm
eff}$ when the characteristic frequencies of the slow modes
measured in DLS are comparable to the frequency at which $T_{\rm
eff}$ is measured. Measuring the frequency dependence of the
effective temperature would be needed to confirm this
interpretation.

The same experimental system has been investigated by the group of
Ciliberto \cite{BellonEPL2001,BellonPhysicaD2002,BuissonJPCM2003}.
They probe the electrical and rheological properties of the
material during the formation of the soft glass. For the
dielectric experiments, the clay solution is used as a conductive
liquid between two electrodes. The set-up allows the frequency
dependence of both the electrical impedance and the voltage noise
to be measured, from which an effective temperature is derived, by
means of a generalized Nyquist formula. They find that $T_{\rm
eff}$ is a decreasing function of frequency and reaches the bath
temperature at high frequency. Moreover, $T_{\rm eff}$ decreases
as the sample ages, while the FDT is strongly violated for young
samples: $T_{\rm eff}$ can be larger than $10^5$ K at low
frequency ($1$ Hz) and small age
\cite{BellonEPL2001,BellonPhysicaD2002}. They show
\cite{BuissonJPCM2003} that the origin of the large violation is
the highly intermittent dynamics characterized by large
fluctuations of the voltage noise (see \sref{sec:hetero} for more
details on the dynamic heterogeneities). We note that, in this
line of thought, Crisanti and Ritort \cite{CrisantiEPL2004} have
shown numerically how an effective temperature can be extracted
from the probability distribution function of intermittent events.
Because of the strong fluctuations of the noise in the experiments
of Ciliberto and coworkers, the value of $T_{\rm eff}$ extracted
from the voltage data is extremely sensitive to rare events and to
the duration over which data are collected. This may explain the
discrepancy between references
\cite{BellonEPL2001,BellonPhysicaD2002} and \cite{BuissonJPCM2003}
concerning the numerical value of the effective temperature
obtained by dielectric measurements (about one order of
magnitude). Bellon and Ciliberto \cite{BellonPhysicaD2002} have
also tested the FDT for the rheological properties of the same
system. To that end, they have built a novel ``zero-applied
stress'' rheometer \cite{BellonRSI2002}, by which the (very small)
strain induced by thermal fluctuations can be measured. Although
strong aging properties are measured in the rheological responses
of the material (as measured also for the dielectric properties),
no violation of the FDT could be detected. The authors propose
several possible explanations for the discrepancy between
dielectric and rheological measurements. In particular, they point
out that the strong fluctuations of the voltage noise could result
from the dissolution of ions in the solution. Ion dissolution has
been very recently confirmed by Bandyopadhyay and coworkers
\cite{BandyopadhyayPRL2004} who argue that the dissociation of the
ions at the surface of the particles leads to the increase of the
interparticle repulsion and is at the origin of the slow dynamics
and aging of Laponite samples. Because the increase of the
interparticle repulsion may lead to a rearrangement event only
when the internal stress thus generated exceeds the (local) yield
stress, fluctuations in the mechanical properties are expected to
be much less strong than fluctuations of the dielectric
properties, which are more sensitive to the release of charges. On
the other hand, the rheology measurements of the group of
Ciliberto are in principle similar (albeit not identical) to those
of Abou and Gallet, since both probe the mechanical properties of
the material. Abou and Gallet measure a weak violation of the FDT
($T_{\rm eff}/T$ is at most equal to $1.8$ at a frequency of $1$
Hz), while the measurements of Ciliberto's group do not detect any
violation. However, it should be pointed out that the rheology
experiments of Ciliberto and coworkers may lack the sensitivity
required to measure violations as small as those reported by Abou
and Gallet. Whether these experimental results are indeed
conflicting remains therefore an open question.

To conclude, we remark that experimental works are still scarce,
owing to the difficulty of measuring simultaneously the response
function and the fluctuation in experiments on soft materials.
Laponite clay suspensions have been mostly investigated and the
results obtained so far by two independent groups and using
different techniques display both qualitative and quantitative
discrepancies. On the other hand, simulations works on a binary
Lennard-Jones mixture \cite{BerthierJChemPhys2002} have shown that
the effective temperature is independent on the chosen observable,
while it has been shown theoretically \cite{FieldingPRL2002} that
the effective temperature does depend on the observable in the
glass phase of Bouchaud's trap model. More theoretical, numerical,
and experimental investigations are needed to rationalize these
contrasting findings.

\section{Dynamical heterogeneity}
\label{sec:hetero}

Dynamical heterogeneity is now recognized as a fundamental feature
of the slow dynamics of supercooled fluids and glasses in hard
condensed matter, thanks to the large body of experimental,
numerical and theoretical work carried out in the last decade (a
general review on dynamical heterogeneity can be found, e.g., in
\rref{RichertJPCM2002}). Most early observations of heterogeneous
dynamics focussed on temporal heterogeneity: the coexistence of
different relaxation times was identified as the source of the
non-exponential relaxations observed in glass formers. Spatial
heterogeneity was often invoked as the most plausible physical
origin of this coexistence (see \cite{RichertJPCM2002} and
references therein). Subsequent experimental and numerical work
has shown that indeed the dynamics of glass formers is spatially
heterogeneous, and spatial heterogeneity has been related to the
cooperative nature of the slow dynamics (for reviews on spatial
heterogeneity that focus on experimental and numerical work, see
for example references \cite{Ediger} and \cite{Glotzer},
respectively).

Cooperativity plays a central role in  many recent theories, where
the glass transition is explained as a dynamical (as opposed to
thermodynamic) transition driven by the divergence of the size of
regions that undergo cooperative rearrangements. Following this
approach, analogies have been drawn with critical phenomena, the
static correlation length of the latter being replaced by a
suitable dynamical correlation length
\cite{GarrahanPNAS2003,WhitelamPRL2004,ChamonJChemPhys2004,ToninelliPRL2004,BiroliEurophysLett2004}.
It should be noted that most theories are developed in the
framework of spin models or the so-called dynamically facilitated
(or kinetically constrained) models \cite{RitortAdvPhys2003},
where the motion of on-lattice particles depends on the number of
occupied neighboring sites. Making quantitative connections
between the results for these systems and molecular glass formers
or colloidal systems may be therefore difficult. However, we note
that recent numerical and theoretical work has shown that the
concept of the glass transition being a dynamical critical
phenomenon can be successfully applied also to Lennard-Jones glass
formers \cite{WhitelamPRL2004} and in the framework of the mode
coupling theory \cite{BiroliEurophysLett2004}.

Experimentally, soft materials provide a unique opportunity to
study in great detail temporal and spatial heterogeneity in
supercooled fluids and glasses, because the relevant length and
time scales are more easily accessible than for hard condensed
matter systems. In this section, we will review recent experiments
that probe dynamical heterogeneity in a variety of systems,
ranging from model hard sphere suspensions to more complicated
glassy samples, such as colloidal gels and concentrated surfactant
phases. Most experiments are performed using time-resolved
confocal scanning microscopy \cite{HabdasCOCIS2002} or recently
introduced light scattering methods that allow temporal
heterogeneities to be measured \cite{LucaJPCM2003}, as it will be
discussed in the following subsections.

\subsection{Optical microscopy experiments}
\label{subsec:heteromicro}

Optical microscopy and digital imaging processing allow one to
follow simultaneously the individual trajectories of a large
number of particles (up to thousands), tracking their position to
an accuracy of a few tens of nanometers
\cite{CrockerJCollInterfaceSci1996}. Three-dimensional motion can
be studied thanks to time-resolved, laser-scanned confocal
microscopy. Once the particle trajectories are known, a wide range
of statistical quantities can be calculated in order to detect,
characterize, and quantify dynamical heterogeneity. Direct
comparison with simulation work is possible, making optical
microscopy a powerful experimental technique for investigating
slow dynamics in glassy soft materials (for a recent review of
video microscopy applied to colloidal suspensions, see
\rref{HabdasCOCIS2002}).

Early optical microscopy measurements of dynamical heterogeneities
were performed by Kasper and coworkers, who studied the self
diffusion of tracer particles in a concentrated hard sphere fluid
\cite{KasperLangmuir1998}. The tracer particles have a core-shell
structure: the outer layer is identical in composition to the host
particles, while the inner core has a large optical contrast with
the fluid and the host particles. The trajectories of the tracer
particles are followed in a thin slice of a three-dimensional
sample, allowing the mean squared displacement (MSD) and the self
part of the van Hove correlation function, $G_{\rm S}(x,t)$, to be
calculated. $G_{\rm S}(x,t)$ represents the (density of)
probability that a particle moves a distance $x$ in a time step
$t$. For a diffusive process, $G_{\rm S}(x,t)$ is a Gaussian
distribution. Kasper $et$ $al.$ find that in concentrated
suspensions $G_{\rm S}(x,t)$ departs from a Gaussian behavior, the
deviations being increasingly marked as $\phi$ approaches $\phig$.
These deviations are due to a small but significant fraction of
displacements $x(t)$ larger than expected. Deviations from
Gaussian behavior are quantified by the so-called non-Gaussian
parameter $\alpha_2$, defined as the fourth moment of $G_{\rm
S}(x,t)$, properly normalized ($\alpha_2 = 0$ for a Gaussian
distribution).

Similar experiments are reported by Kegel and van Blaaderen
\cite{KegelScience2000}, who use confocal microscopy to study in
two dimensions the dynamics of concentrated suspensions of nearly
density- and refractive index-matched colloids behaving as hard
spheres. In contrast to the experiment of Kasper $et$ $al.$, all
particles are tracked (as opposed to tracer particles only),
improving dramatically the statistics. For concentrated
suspensions in the fluid phase, $G_{\rm S}(x,t)$ is reasonably
well described by the sum of two Gaussian distributions,
corresponding to two distinct populations of ``fast'' and ``slow''
particles. This result clearly demonstrate the heterogenous nature
of the dynamics, in agreement with several simulation works (see,
e.g., reference \cite{KobPRL1997}).

Weeks and coworkers performed the first three-dimensional optical
microscopy investigation of the slow dynamics of supercooled
colloidal suspensions \cite{WeeksScience2000}. They use PMMA
particles stained with a fluorescent dye and suspended in an
organic solvent that nearly matches both the density and the
refractive index of the particles. Note that under the reported
experimental conditions the particles are slightly charged
\cite{YethirajNature2003}. They find a non-Gaussian $G_{\rm
S}(x,t)$, with a non-Gaussian parameter that has a peaked shape as
a function of the time step $t$. The peak position corresponds to
the characteristic time of the $\alpha$ relaxation, and its height
increases when $\phi$ approaches $\phig$, indicating increasingly
heterogeneous behavior. In the glass phase, no clear peak is
observed. In analogy with previous simulation work
\cite{KobPRL1997,DonatiPRL1998}, Weeks and coworkers study the
spatial arrangement of the most mobile particles (the mobility is
measured for time intervals comparable to the $\alpha$ relaxation
time). They find that these particles form clusters that have a
fractal morphology (fractal dimension $\approx 1.9$) and whose
size, for supercooled suspensions, increases with increasing
$\phi$, up to a radius of gyration of about 10 particle radii. For
glasses, defining the most mobile particles is more difficult,
since no $\alpha$ relaxation is observed and the cluster size
depends sensitively on the way the mobility is calculated
\cite{CourtlandJPCM2003}. When the particle trajectories are
slightly averaged over time to remove the contribution of Brownian
motion, the clusters observed in glasses are similar to those
observed for supercooled samples. Interestingly, no change in the
cluster size is observed for glasses during aging, thus ruling out
the increase of the correlation length of the dynamics as a
possible origin for the slowing down of the dynamics
\cite{CourtlandJPCM2003}.

In glassy systems, the relationship between dynamical
heterogeneity and local structure is a long-standing open
question. Various experiments suggest that particle mobility is
related to the degree of local disorder and packing, although it
should be stressed that very small changes in the local structure
are associated to huge variations in the dynamics. In the same
three-dimensional experiment discussed above, Weeks and Weitz have
shown that the clusters of the most mobile particles occupy
regions with lower local density and higher disorder
\cite{WeeksPRL2002}. Two-dimensional systems allow a more direct
investigation of the relationship between local structure and
dynamical heterogeneity. Cui and coworkers have studied the
dynamics of a single layer of concentrated monodisperse particles
confined in a quasi-two-dimensional cell \cite{CuiJChemPhys2001}
(note that this system does not exhibit a glass phase, but rather
crystallizes at high enough volume fraction). At relatively high
volume fractions and intermediate time scales, the dynamics is
very heterogeneous, with two distinct populations of particles.
The fast particles move in a string-like fashion ---strongly
reminiscent of that observed in simulations of supercooled fluids
\cite{DonatiPRL1998}--- along channels formed by the disordered
boundaries between regions of slowly-moving, quasi-ordered
particles. On very long time scales, infrequent, large
displacements are associated with transient regions of lower
density created by density fluctuations. A very elegant
realization of a two-dimensional colloidal glass former has been
recently reported by K\"{o}nig and coworkers
\cite{KonigAIP2004,KonigPRL2004}. Superparamagnetic colloidal
particles are confined at the water-air interface of a hanging
liquid drop; the interactions between particles can be fine-tuned
by applying a magnetic field, thereby fixing the ``temperature''
of the system \cite{ZahnPRL1997}. Mixtures of particles of two
different sizes are used, in order to prevent crystallization. The
structure is analyzed in terms of locally ordered patterns of
small and large particles that maximize the packing density
\cite{KonigAIP2004bis,KonigPRL2004bis}. Dynamical heterogeneities
appear to be associated with the regions where the local order is
frustrated and the local packing is looser
\cite{KonigAIP2004,KonigAIP2004bis}.

\subsection{Light scattering experiments}
\label{subsec:heteroscattering}

Although in light scattering experiments no microscopic
information on the individual particle trajectories are available,
this technique still presents some attractive features compared to
optical microscopy. The scattering volume is typically larger than
the sample volume imaged by microscopy, thus allowing a better
statistics to be achieved. A large experimental volume may be
particularly important when the dynamics is spatially correlated:
for example, for samples near $\phig$, the size of the clusters of
fast-moving particles measured in the experiments by Weeks $et$
$al.$ \cite{WeeksScience2000} was comparable to the full field of
view, thus making difficult their precise characterization.
Moreover, particles used in light scattering experiments are
typically smaller than those used for imaging: as a consequence,
the time scales of the slow dynamics are not as prohibitively long
as for larger particles. In addition, small particles are less
influenced by external fields such as gravity. Finally,
experiments can be performed both in the single and in the
strongly multiple scattering limits, thus extending the possible
choice of systems (by contrast, optical microscopy requires nearly
index-matched suspensions, except for two-dimensional systems).

Traditional light scattering techniques, however, do not provide
direct information on dynamical heterogeneity, because of space
and time averaging. In fact, the intensity correlation function
$g_2 -1$ has to be averaged over extended periods of time in order
to achieve an acceptable accuracy (typically up to four orders of
magnitudes longer than the largest relaxation time in the system),
and the detector collects light scattered by the whole illuminated
sample. Indirect measurements of dynamical heterogeneities are
still possible: for example, the non-Gaussian parameter $\alpha_2$
that quantifies deviations from diffusive behavior can be measured
(see e.g. reference \cite{VanMegenJChemPhys1988}). However, it
should be noted that a non-zero value of $\alpha_2$ could be due
either to the coexistence of fast and slow populations of
particles, as discussed in the previous subsection, or to the same
non-diffusive behavior shared by all particles. Traditional light
scattering experiments lack the capability of discriminating
between these two contrasting scenarios.

Higher-order intensity correlation functions contain more
information on temporal heterogeneities of the dynamics
\cite{LemieuxJOSAA1999,LemieuxAPPOPT2001}. In particular, Lemieux
and Durian have studied the dynamics of the upper layer of grains
in a heap upon which grains are steadily poured at a flow rate $Q$
\cite{LemieuxPRL2000}. They measure the 4-th order intensity
correlation function, $g^{(4)}_T(\tau) = \langle
I(t)I(t+T)I(t+\tau)I(t+\tau+T)\rangle_t / \langle I \rangle_t ^4$,
where $I(t)$ is the multiply scattered intensity measured by a
point-like detector and the average is over time $t$\footnote{In
this section we follow the notation of most experimental works by
indicating time by $t$ and a time lag by $\tau$. This is different
from the notation used in previous sections of this paper and in
most numerical and theoretical works where $t$ and $t'-t$ are
used, respectively.}. They show that, contrary to $g_2$, $g^{(4)}$
allows intermittent, avalanche-like processes (occurring at low
$Q$) to be distinguished from continuous dynamics (occurring at
high $Q$). Moreover, both the statistics of the avalanches and the
motion of grains within a single avalanche can be obtained by
analyzing $g^{(4)}$.

The technique proposed by Lemieux and Durian still requires to
average the (higher-order) intensity correlation functions over
time. This poses a problem when studying very slow or
non-stationary processes, as it is often the case for soft
glasses. Cipelletti and coworkers have recently proposed an
alternative approach, termed Time Resolved Correlation (TRC)
\cite{LucaJPCM2003,DuriFNL2004}, where one takes full advantage of
the multispeckle method. In a TRC experiment, a CCD camera is used
to record a time series of pictures of the speckle pattern of the
light scattered by the sample (both single scattering and DWS
measurements are possible). The dynamics is quantified via $\cI$,
the instantaneous degree of correlation between pairs of speckle
patterns recorded at time $t$ and $t+\tau$: $\cI = \frac{\langle
I_p(t)I_p(t+\tau) \rangle_p}{\langle I_p(t) \rangle_p \langle
I_p(t+\tau) \rangle_p} -1$. Here, $I_p$ is the intensity measured
by the $p$-th pixel and averages are over all CCD pixels. Because
any change in the sample configuration results in a change in the
speckle pattern, $\cI$ quantifies the overlap between sample
configurations separated by a time lag $\tau$, as a function of
$t$. The time average of $\cI$ yields the intensity correlation
function $g_2(\tau)-1$ usually measured in light scattering, while
the raw $\cI$ is analogous to the two-time correlation function
studied numerically or experimentally for non-stationary (e.g.
aging) systems. Note however that in most simulations and
experiments the two-time correlation function is averaged over a
short time window or over different realizations of the system, in
order to reduce its ``noise''. By contrast, the essence of the TRC
method is to extract useful information from the fluctuations of
$c_I$.

\begin{figure}
\begin{center}
\includegraphics[scale=0.8]{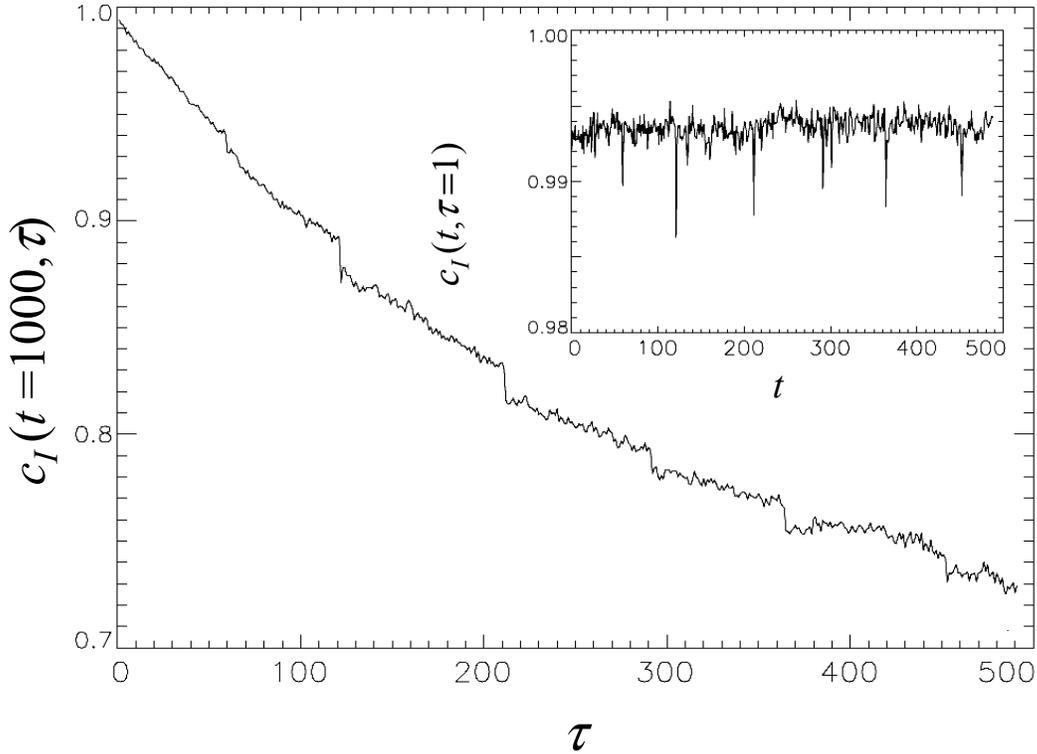}
\end{center}
\caption{\label{fig:cI} Time Resolved Correlation measurements for
a granular material gently tapped. The dynamics of this athermal
system is due to the taps; therefore both the time, $t$, and the
time delay, $\tau$, are expressed in number of taps. Inset: $\cI$
as a function of $t$ for a fixed time delay $\tau = 1$. Note the
rare, large drops of the TRC signal, indicative of intermittent
dynamics. Main plot: for the same system, $\cI$ is plotted as a
function of time delay, for fixed $t=1000$. The step-like
relaxation of $c_I$ is due to large, rare rearrangement events
similar to those that yield the downward spikes in the inset.
Adapted from \rref{KablaPhD2003}.}
\end{figure}

In order to investigate the temporal heterogeneity of the
dynamics, it is useful to plot $\cI$ as a function of time $t$,
for a fixed lag $\tau$. For temporally homogeneous dynamics, one
expects $\cI$ to be constant (except for small fluctuations due to
measurement noise), as verified on dilute suspensions of Brownian
particles \cite{LucaJPCM2003,DuriFNL2004}. On the contrary, a
large drop of $c_I$ at time $t$ would be indicative of a sudden
rearrangement event occurring between $t$ and $t+\tau$ and leading
to a significant change of the sample configuration. Large drops
of $c_I$ have been indeed observed in TRC measurements on a
variety of systems, including colloidal fractal gels and
concentrated surfactant phases
\cite{LucaJPCM2003,BissigPhysChemComm2003}, flocculated
concentrated colloidal suspensions \cite{Sarcia2004}, and granular
materials \cite{KablaPRL2004,KablaPhD2003,CaballeroCondMat2004}.
As an example, the inset of \fref{fig:cI} shows TRC data for the
granular material studied in \rsref{KablaPRL2004,KablaPhD2003}:
large drops of the degree of correlation are clearly visible, thus
demonstrating the intermittent nature of the slow dynamics. These
sudden rearrangements are also visible in the two-time correlation
function, provided that no average is performed, as discussed
above. The main plot of \fref{fig:cI} illustrates this point by
showing $\cI$ as a function of $\tau$ for a fixed $t$, for the
same system as in the inset. The rearrangement events result in a
discontinuous, step-like relaxation of the correlation function
(note that a similar behavior was observed in simulations on a
molecular glass former, when the two-time correlation function was
not averaged over different system realizations
\cite{KobEPJB2000}). By contrast, the $average$ intensity
correlation function $g_2 -1$ is often indistinguishable from that
of a system with homogeneous dynamics. This demonstrates that
great care should be used in applying standard equilibrium methods
to extract from $g_2$ quantities such as the particles' MSD,
because these methods usually assume the dynamics to be
homogeneous. In particular, the preliminary experiments reported
in reference \cite{LucaJPCM2003} on the same systems for which the
``ballistic'' motion associated with internal stress relaxation
was observed (see sections \ref{subsec:slow} and
\ref{subsec:internalstress}) raise the issue of how to reconcile
intermittent dynamics with ballistic motion. A simple explanation
would be to assume that the dynamics is due to a series of
individual rearrangements and that the particle motion
---in a given region of the sample--- resulting from distinct
rearrangements is highly correlated (uncorrelated motion would
lead to a diffusive-like behavior). The model proposed by Bouchaud
and Pitard for the slow dynamics of colloidal fractal gels is
indeed based on this idea \cite{BouchaudEPJE2001}.

An important point in the TRC experiments reported above is that
the technique allows one to probe directly temporal heterogeneity,
not spatial heterogeneity, because each CCD pixel receives light
issued from the whole scattering volume. However, the very
existence of large temporal fluctuations of $\cI$ demonstrates
---albeit indirectly--- that the dynamics is spatially correlated
over large distances. More precisely, in order to observe
intermittent dynamics the number $N_d$ of dynamically independent
regions contained in the scattering volume has to be limited,
because for $N_d \rightarrow \infty$ many rearrangement events
would occur in the sample at any given time, leading to small,
Gaussian fluctuations of $\cI$. Similar arguments have been
invoked in recent simulation and theoretical works to explain
non-Gaussian fluctuations in systems with extended spatial or
temporal correlations (see for example references
\cite{CrisantiEPL2004,ChamonJChemPhys2004,PitardCondMat2004,BramwellNature1998,BramwellPRL2000,CluselPRE2004,CastilloPRL2002,
SibaniCondMat2004}). In particular, the probability density
function (PDF) of $c_I$ at fixed $\tau$ is in many cases strongly
reminiscent of the ``universal'' Gumbel distribution
\cite{BramwellNature1998,BramwellPRL2000} or of a generalized
Gumbel-like PDF \cite{ChamonJChemPhys2004}. These distributions
are characterized by an asymmetric shape, with an (asymptotically)
exponential tail. We note however that the detailed shape of the
PDF of $c_I$ depends sensitively on the lag $\tau$
\cite{BissigPhDThesis}, a feature also found in theoretical work
\cite{PitardCondMat2004}.

In the TRC experiments discussed above, individual rearrangement
events can be identified and the probability distribution of the
time interval, $t_e$, between such events can be calculated. For
the flocculated suspension studied in \rref{Sarcia2004} Sarcia and
H\'{e}braud find a power law distribution $t_e \sim t^{-2}$.
Interestingly, this behavior agrees with theoretical predictions
derived in the framework of the trap model for glassy systems, in
the regime where a limited number of dynamically independent
regions are observed \cite{PitardCondMat2004}. For other
experimental systems, individual rearrangement events may not be
distinguishable, because the drop of $c_I$ associated with one
single event may be negligible. In this case, a measurable loss of
correlation is always due to the cumulative effect of many events,
even for the smallest delays $\tau$ accessible experimentally.
Deviations from temporally homogeneous dynamics are still
detectable, however. An example is provided by the dynamics of a
shaving cream foam \cite{MayerPRL2004}: $\cI$ is found to exhibit
large fluctuations on a time scale much longer than the average
relaxation time of the intensity autocorrelation function. As a
consequence, the two-time correlation function decays smoothly at
any time (no step-like relaxation such as that in \fref{fig:cI}b
is observed), but its relaxation time slowly fluctuates with time
$t$. These fluctuations are quantified by introducing the variance
of $c_I$, defined by $\chi(\tau) = \langle \cI^2 \rangle_t -
\langle \cI \rangle^{2}_t$. $\chi(\tau)$ is the analogous for TRC
experiments of the generalized dynamical susceptibility  $\chi_4$
introduced in theoretical and numerical works on spin systems,
hard spheres, and molecular glass formers
\cite{MayerPRL2004,DonatiCondMat1999,LacevicJChemPhys2003}. Note
that $\chi_4$ is proportional to the volume integral of the
so-called four-point density correlation function, which compares
the change of the local configuration around the position
$\textbf{r}_1$ during a time interval $\tau$ to the corresponding
quantity for position $\textbf{r}_2$
\cite{MayerPRL2004,LacevicJChemPhys2003}. Because $\chi_4$ is
related to the spatial correlations of the dynamics, this
parameter provides a quantitative link between temporal and
spatial dynamical heterogeneity. For the foam, $\chi(\tau)$ has a
peaked shape, the largest fluctuations of the dynamics being
observed on a time scale comparable to $\tau_s$, the average
relaxation time of the system. This feature is strongly
reminiscent of the behavior of the systems investigated
numerically and theoretically in
\rsref{PitardCondMat2004,MayerPRL2004,LacevicJChemPhys2003}.

In simulations of glass formers, the height of the peak of
$\chi_4$ increases when approaching the glass transition
\cite{LacevicJChemPhys2003}. This behavior has been interpreted as
due to the increase of the size of dynamically correlated regions.
The foam and the Ising spin models studied in \rref{MayerPRL2004}
provide a means to test this hypothesis on systems for which there
is a natural characteristic length that increases with time, due
to coarsening (this length is the bubble size and the size of
parallel spin domains for the foam and the Ising models,
respectively). Indeed, for both systems the same dynamic scaling
of fluctuations with domain size is observed: $\chi(\tau / \tau_s)
\sim N^{-1}$, where $\tau_s$ is the average relaxation time of the
correlation function and $N$ is the number of bubbles (or spin
domains) contained in the system. Similar TRC experiments have
been performed on a very polydisperse colloidal paste by Ballesta
and coworkers \cite{BallestaAIP2004}. They find that $\chi(\tau)$
has a peaked shape, similarly to the systems reported above.
Moreover, the height $\chi^*$ of the peak evolves in a very
surprising way: initially $\chi^*$ increases with increasing
volume fraction $\phi$, similarly to the growth of the peak in
supercooled molecular systems upon cooling
\cite{LacevicJChemPhys2003}. In contrast with molecular systems,
however, $\chi^*$ reaches a maximum value at intermediate volume
fractions and is drastically reduced at the highest $\phi$
experimentally achievable \cite{BallestaUnpublished2004}. More
experiments on different systems will be needed to test the
generality of this behavior. In particular, model systems such as
hard spheres suspensions are an attracting choice, since detailed
predictions on the behavior of $\chi(\tau)$ in the framework of
the MCT are now available \cite{BiroliEurophysLett2004}.

\subsection{Other experiments and concluding remarks on dynamical heterogeneity}
\label{subsec:heteroother}

Dynamical heterogeneity has been measured also by other techniques
and in different systems. Strongly intermittent behavior has been
observed in the dielectric signal for a Laponite glass
\cite{BuissonJPCM2003}, as already mentioned in section
\ref{subsec:FDT}. As a result, the PDF of the voltage signal
measured in these experiments exhibits a non-Gaussian behavior,
with roughly exponential tails similar to those reported in
reference \cite{CrisantiEPL2004}. The microscopic origin of these
events is still unclear. Attempts to measure intermittent behavior
in the rheological response of the same Laponite system using the
``zero-shear'' rheometer \cite{BellonRSI2002} mentioned in
\sref{subsec:FDT} were, to date,  not successful.

Interestingly, heterogeneous dynamics has been reported for a
variety of granular systems, using different techniques.
Intermittency has been measured by TRC in granular systems gently
vibrated \cite{KablaPRL2004,KablaPhD2003,CaballeroCondMat2004}.
These experiments show that the evolution of the system
configuration is not continuous, but rather occurs through a
series of discrete rearrangement events, similar to those reported
in \rsref{LucaJPCM2003,BissigPhysChemComm2003,Sarcia2004} for
colloidal systems. On a more microscopic level, Marty and Dauchot
\cite{MartyCondMat2004} and Pouliquen and coworkers
\cite{PouliquenPRL2003} have tracked the motion of individual
``grains'' in two-dimensional and three-dimensional granular
systems submitted to a cyclic shear. They find that the grain
motion is strongly reminiscent of that of colloidal glasses: a
cage effect is observed and deviations from a Gaussian behavior in
the PDF of the grain displacement are reported, similarly to the
results of the experiments discussed in \sref{subsec:heteromicro}.
Collectively, these analogies support and extend to a microscopic
level the unifying picture underlying the concept of jamming
proposed by Liu and Nagel \cite{LiuNature1998}.

As a final remark, we note that intriguing analogies exist also
between the heterogeneity of the spontaneous dynamics of soft
glasses and the behavior of many complex fluids under shear.
Indeed, for concentrated colloidal suspensions and surfactant
systems the macroscopic rheological response to an applied
constant strain or stress has been shown to exhibit large temporal
fluctuations
\cite{LootensPRL2003,SalmonPRE2002,SalmonPRE2003II,CourbinPRL2004}.
On the other hand, simulations and experiments have shown that the
stress relaxation and the flow of foams, ultrasoft glasses and
concentrated surfactant systems  are both temporally and spatially
heterogeneous (see e.g. references
\cite{SalmonPRE2003II,SalmonPRE2003I,DebregeasPRL2001,KablaPRL2003,BecuPRL2004,HolmesJRheol2004}).
Although this heterogeneous behavior is usually associated to the
existence of different microstructures (e.g. disordered/ordered
onion phases in \cite{SalmonPRE2003I}), simulations of a simple
model of an (athermal) generic yield stress fluid suggest that
heterogeneity may exist even in structurally homogeneous fluids
\cite{PicardCondMat2004}. In these simulations, the flow of a
yield stress fluid sheared at a constant rate is shown to be due
to discrete rearrangement events, whose characteristic size
diverges for a vanishing shear rate. The basic ingredients of this
model (localized plastic events due to a microscopic yield stress
and elastic propagation of the local stress relaxation) are
certainly relevant for many soft glasses. In particular, close
analogies may be drawn between slowly sheared yield stress systems
and elastic glassy materials
---such as colloidal gels--- where internal stresses are
progressively built up and induce particle rearrangements. More
work will be needed to explore in depth these analogies and to
fully assess their relevance and generality.

\section{Conclusion}
\label{sec:conclusions}

In this paper, we have reviewed the experimental work of the last
few years on the slow dynamics in soft matter. In particular,
research in four areas has been discussed: the existence of two
different glass states (attractive and repulsive), the dynamics
and the aging  of systems far from equilibrium, the effect of an
external perturbation on glassy materials, and dynamical
heterogeneity. As it was pointed out in many occasions in the
preceding sections, these topics are closely related and indeed
the same systems have been often studied in the context of two or
more of these areas. At the end of each section we have listed
what are, in our opinion, the most relevant questions that are
still open: here, we limit ourselves to some brief final
considerations.

A great effort has been directed to developing unified approaches
that may account for the similarities in the slow dynamics of many
soft materials and for the fascinating analogies with hard
condensed matter glasses and granular media. The most successful
theory is probably the MCT: initially, it was restricted to hard
sphere suspensions, but it has now been extended to include
attractive systems at moderate to high volume fractions. However,
it should be noted that the deep physical reasons of its success
are still somehow unclear; for example, dynamical heterogeneity,
now recognized as a key feature of the slow dynamics of most
glassy systems, is not included in standard formulations of the
MCT. Very recent theoretical work shows that quantitative
predictions on dynamical heterogeneity can be made in the
framework of the MCT: experiments that test these predictions
would certainly shed new light on the limits to the validity of
the MCT.

An alternative unifying approach is the jamming scenario, which is
however a conceptual tool rather than a fully-developed
quantitative theory. Its appeal resides in the large variety of
systems whose fluid-to-solid transition may be rationalized in a
unified picture, ranging from colloidal suspensions to molecular
glasses and granular materials. The analogies in the local
dynamics between glass formers and granular media that have been
rapidly reviewed in this paper (cage effect, dynamical
heterogeneity) provide additional support to the jamming scenario.
Another intriguing similarity hinted to by recent work is the
increase of heterogeneity in unperturbed glassy systems when
temperature or packing fraction approaches a critical value, which
is paralleled by the growth of spatial and temporal heterogeneity
in yield-stress fluids when the shear rate vanishes. This analogy
is particularly suggestive in the framework of jamming, because of
the similar role that is attributed to temperature (or
interparticle potential and volume fraction) and stress in driving
the fluid-to-solid transition.

While the MCT and the jamming scenario provide some guidance in
describing systems that approach the non-ergodicity transition
from the fluid phase, our understanding of materials deeply
quenched in an out-of-equilibrium phase is less advanced. Although
some common features are observed (dynamical heterogeneity and a
general trend for the dynamics to slow down with sample age being
two of the most prominent), a large palette of different behaviors
is observed, which remains largely unexplained (see, e.g., the
various aging regimes discussed in this review). The concept of
effective temperature may prove useful to describe quantitatively
out-of-equilibrium soft materials and their aging; however,
current experimental determinations of $T_{\rm{eff}}$ are still
too scarce and yield contradictory results. In particular,
simultaneous measurements of the effective temperature for various
observables and in a wider variety of systems will be necessary.
Finally, we observe that stress relaxation appears to be an
important ingredient of the slow dynamics of many soft glasses; in
spite of that, it has been generally neglected in theoretical
approaches, with the exception of the model for colloidal gels
cited in section \ref{subsec:slow}.

Recent advances in the investigation of slow dynamics in glassy
soft matter have been made possible by close interactions between
theory, simulation and experiments. On the experimental side, new
methods and techniques have been developed to extend measurements
to out-of-equilibrium and very slowly relaxing systems. A great
effort is currently done to obtain spatially- and time-resolved
information on the dynamics, and to identify the most insightful
quantities to be extracted from the raw data in order to
characterize heterogeneous behavior. Future advances will likely
include the combined (and possibly simultaneous) use of these
methods, in order to achieve a more complete understanding of the
various physical mechanisms that drive the relaxation of glassy
soft matter, as well as of their interplay. Conceptually, the
focus will be in identifying and explaining  the ``universal''
features of the slow dynamics in vastly different systems.

\ack

We thank the numerous colleagues that shared with us their results
before publication. E. Zaccarelli, V. Viasnoff, and A. Kabla
kindly provided us with their data for some of the figures of this
review: we thank them warmly. We are indebted to many people for
very useful and stimulating discussions; in particular we wish to
thank L. Berthier, E. Pitard, and V. Trappe. We are grateful to
present and past coworkers in our group: P. Ballesta, A. Duri, and
S. Mazoyer. Financial support for our research on slow dynamics
was provided by the French Minist\`{e}re pour la Recherche (ACI
JC2076), R\'{e}gion Languedoc-Roussillon, CNRS (PICS 2410 and
Projet mi-lourd ``Dynamiques lentes et vieillissement de
mat\'{e}riaux''), CNES (grants no. 02/4800000063 and
03/4800000123), and the European Community through the
``Softcomp'' Network of Excellence and the Marie Curie Research
and Training Network ``Arrested Matter''(grant no.
MRTN-CT-2003-504712). L. C. thanks the Institut Universitaire de
France for supporting his research.

\end{document}